\documentclass{aa}  
\usepackage[varg]{txfonts}
\usepackage{graphicx}
\usepackage[colorlinks=true,linkcolor=blue,citecolor=blue, urlcolor=blue]{hyperref}

\begin{document}

\title{Polarimetry of the Ly$\alpha$ envelope of the radio-quiet quasar SDSS~J124020.91+145535.6
\thanks{Based on
observations made with the  FORS2 imager mounted on
the Antu  VLT   telescope  at   ESO-Paranal   Observatory   (programme 
092.B-0448; PI: P. North)}}
\author{P.~North\inst{\ref{inst1}} \and M.~Hayes\inst{\ref{inst2}}
\and M.~Millon\inst{\ref{inst3}}
\and A.~Verhamme\inst{\ref{inst4}}
\and M.~Trebitsch\inst{\ref{inst5}}
\and J.~Blaizot\inst{\ref{inst6}} \and
F.~Courbin\inst{\ref{inst1}}
\and D.~Chelouche\inst{\ref{inst7}}}
\institute{Institut de Physique, Laboratoire d'astrophysique, 
           Ecole Polytechnique F\'ed\'erale de Lausanne (EPFL),
           Observatoire de Sauverny,
           CH-1290 Versoix, Switzerland\label{inst1}
\and
           Stockholm University, Albanova University Center,
           Department of Astronomy,
           SE-106 91 Stockholm, Sweden\label{inst2}
\and
           Kavli Institute for Particle Astrophysics and Cosmology and Department of Physics, Stanford University, Stanford, CA 94305, USA \label{inst3}
\and
           Observatoire de Sauverny, Universit\'e de Gen\`eve,
           Chemin Pegasi 51,
           1290 Sauverny, Switzerland\label{inst4}
\and
           Kapteyn Astronomical Institute,
           University of Groningen, P.O. Box 800,
           9700AV Groningen, The Netherlands\label{inst5}
\and
           Univ. Lyon, Univ. Lyon 1, ENS de Lyon, CNRS, Centre de Recherche Astrophysique de Lyon
           UMR5574,
           69230 Saint-Genis-Laval, France\label{inst6}
\and
           Department of Physics, Faculty of Natural Sciences,
           University of Haifa, Haifa 31905, Israel\label{inst7}}
\date{Received 11 July 2023/ Accepted 5 January 2024}
\abstract{The radio quiet quasar SDSS~J1240+1455 lies at a redshift of $z=3.11$, is surrounded
by a Ly$\alpha$ blob (LAB), and is absorbed by a proximate damped Ly$\alpha$ system. In order to better
define the morphology of the blob and determine its emission mechanism, we
gathered deep narrow-band images isolating the Ly$\alpha$ line of this object in
linearly polarized light. We
provide a deep intensity image of the blob, showing a filamentary structure
extending up to $16$" (or $~122$~physical~kpc)
in diameter. No significant polarization
signal could be extracted from the data, but $95$\% probability upper limits were defined
through simulations. They vary between $\sim 3$\% in the central $0.75$\arcsec\ disk
(after subtraction of the unpolarized quasar continuum) and $\sim 10$\% in the
$3.8-5.5$" annulus. The low polarization suggests that the Ly$\alpha$ photons are emitted
mostly {\sl \emph{in situ}}, by recombination and de-excitation in a gas largely ionized by
the quasar ultraviolet light, rather than by a central source and scattered subsequently
by neutral hydrogen gas.
This blob shows no detectable polarization signal, contrary to LAB1, a brighter and more
extended blob that is not related to the nearby active galactic nucleus (AGN)
in any obvious way, and where
a significant polarization signal of about $18$\% was detected.}


\titlerunning{Polarimetry of the Ly$\alpha$ blob J1240+1455}

\maketitle
\authorrunning{North et al.}

\section{Introduction}
Diffuse gas around quasars and galaxies bears the testimony of several processes related to galaxy formation and evolution: outflows, due to, for instance, quasar feedback \citep{CMG15}, accretion of intergalactic gas \citep{LHH21}, or tidal debris resulting from collisions \citep{XCA22}.
Giant Lyman-$\alpha$ nebulae or Lyman-$\alpha$ blobs (hereafter LABs) have been
discovered at high redshift over the last three decades, either in isolation or around radio-loud and radio-quiet
quasars \citep{H91a,bremer92,HME96,LB98, MWF98,FMW99,BSS03,VM03,MYHT04,WMF05,CJW06,FMD06,CLP07,
SJ07,SJS09,MNM09,HPKZ09,ZDD09,SCT09,YZTE09,RBH11,MYH11,WCB11,ZML11,NCE12,GUW12,FDC13,
RBH13,KMM14,CAP14,MCM14a,MCM14b,MMM15,JZP16,FPN16,BCL16}.
They typically extend over tens of kiloparsecs, sometimes over more than $100$~kpc.

The origin of the 
emission of these objects has been much debated.
Some authors \citep[e.g.,][]{CKR16,GNM16} attribute it to star formation and resonant
scattering of Ly$\alpha$ photons. Others propose that photoionization by a
central AGN is the main cause \citep{HR01}, even though an AGN is not detected at first
sight in many LABs \citep{GALS09,OND13,SML16,KST17}.
\citet{CCZ16} argued that the emission is due to collisional excitation of gas
that is shock-heated by supernovae and gravitational accretion.
Cooling radiation is also invoked, where cool streams of gas falling into the
potential well of the forming galaxy release gravitational binding energy, which
finally excites the hydrogen atoms \citep{HSQ00,FKG01,NFM06,DL09,RB12,SMB17}.
\citet{DRV22} found some evidence for that.
On the other hand, a Ly$\alpha$ nebula that was long considered as the most convincing
example of gravitational cooling \citep{NFM06} is probably illuminated by a nearby buried
AGN \citep{PMB15, SPC21}.

Linear Ly$\alpha$ polarization was predicted as a diagnostic to infer the nature
of the Ly$\alpha$ emission in LABs; scattered emission may be polarized (depending
on the \ion{H}{i} column density and number of scattering events), while
{\sl \emph{in situ}} emission is not \citep{DL08}. These theoretical predictions were refined
in the context of gravitational cooling Ly$\alpha$ production, where simulations
by \citet{TVB16} showed that gravitational cooling conserves some polarization in
the Ly$\alpha$ nebulae. This is because, even though most Ly$\alpha$  photons are
produced {\sl \emph{in situ}}, they are emitted mostly in the central, densest region
of the nebula, so that a non-negligible part of them reach the outskirts of the nebula
where they are scattered onto the observer.

The first attempt at detecting the polarization in a LAB (object LABd05 discovered by \citet{DBS05}, which contains an obscured AGN) could only put an upper limit of a few percent on the polarization fraction \citep{PSS11}. Using a larger telescope (the MMT-6.5m instead of the Bok-2.3m) and a more sophisticated imaging technique, \citet{KYZ20} later found a clear polarization signal in the same object.
\citet{HSS11} managed to detect significant polarization in the bright
LAB1 blob in the SSA22 protocluster \citep{SAS00} on the basis of imaging
polarimetry. Using the same technique, \citet{YZS17} also detected significant polarization along the radio lobes of a radio galaxy.
These results suggest that
Ly$\alpha$ photons are not produced {\sl \emph{in situ}}, but rather
near the center of the nebula, and scattered by neutral hydrogen
towards the observer. Interestingly, \citet{GBA14} observed an Ultra Luminous InfraRed Galaxy (ULIRG) with the James Clerk Maxwell Telescope
(JCMT) at the center of the polarized pattern and concluded that this galaxy may
well be the source of the Ly$\alpha$ photons scattered by the nebula. It was
subsequently resolved with the Atacama Large Millimeter/Submillimeter
Array (ALMA) into three components \citep{GNM16}, and the main conclusion remains,
that active star formation in these galaxies provides the necessary amount of
Ly$\alpha$ photons to explain the nebula emission.
\citet{CBS05} also showed that SubMillimeter Galaxies (SMGs; which are dusty starbursts similar to ULIRGs) often display Ly$\alpha$ in emission.
The polarization fraction $P$ is maximum for a scattering
angle of $\pi/2$, so the observed fraction is expected to increase with the
projected distance from the LAB center, and this is indeed observed, with $P$
reaching $\sim 20$\% at a radius of $6-7$\arcsec. Furthermore, the polarization
vector is expected to be oriented tangentially (i.e., in a direction perpendicular to that of the source of Ly$\alpha$ photons), which is also observed.
\citet{BSH16} confirmed these results using spectropolarimetry and found
indication of outflow.

Significant polarization ($P=16.4\pm 4.6$\%) was found in part of an extended
Ly$\alpha$ nebula surrounding a radio galaxy \citep{HVV13}, showing yet another
example where scattering of Ly$\alpha$ photons by neutral hydrogen is important.
Similarly, \citet{YZS17} find, in a blob surrounding a radio-loud AGN, $P<2$\%
at $\sim 5$~kpc from the AGN and $P\simeq 17$\% $15-25$~kpc from it; polarization
is detected along the major axis of the blob.
Most recently, \citet{KYZ23} reported the detection of significant polarization in three out of four LABs hosting an AGN.

In this work, we focused on the LAB surrounding the radio-quiet quasar (RQQ)
SDSS~J124020.91+145535.6. As shown by \citet{HPKZ09}, this object is associated
with a proximate damped Ly$\alpha$ system (PDLA), which acts as a natural coronagraph
suppressing the quasar light around the redshifted Ly$\alpha$ wavelength.
Such a providential geometry eases the observation of the faint LAB. As the quasar
provides an obvious source of  Ly$\alpha$ photons that may be scattered by the
hydrogen atoms of the LAB, SDSS~J124020.91+145535.6 appeared as a natural target
for imaging polarimetry through a narrow band (NB) filter centered on the Ly$\alpha$
line. This is the first attempt at observing the polarization of a LAB centered
on a quasar that is not radio loud.\footnote{Some radio-quiet quasars have become radio-loud within a decade or two \citep{NDP20}. This shows that the border between the two categories is not as tight as once thought.  On the basis of the available data, we cannot guarantee that our object was still radio-quiet at the time of observations,  but such transition cases seem relatively rare.}

The observations are
described in Section \ref{observations} and the results are presented in
Sections \ref{sec:results1} and \ref{sec:polar}. The results are discussed assuming a cosmology with
$H_0=70$~km\,s$^{-1}$\,Mpc$^{-1}$, $\Omega_M=0.3,$ and $\Omega_\Lambda =0.7$.
The magnitudes are reported in the Vega photometric system.

\section{Observations and their reduction}
\label{obs-redu}

\subsection{Observations}
\label{observations}
The  observations were carried  out on 1-7 March, 2014 with the
FOcal Reducer and low dispersion Spectrograph (FORS2) instrument \citep{AFF98}
attached to  VLT-UT1.  The FORS2 instrument was used in the imaging
polarimetric (IPOL) mode: after passing through the collimator, the light
beam goes through a Wollaston prism where it is split into two beams
(the ordinary and extraordinary ones) with orthogonal polarizations. For
the two images to be visible without overlapping one another, a strip mask is
placed before the Wollaston prism, removing half the field of view, the full extent of which is $6.8\arcmin\times 6.8\arcmin$. The
strips are 22\,\arcsec\ wide, which is more than the expected size of our
object. Then, the beam goes through a half wave plate (HWP)
that rotates the plane of polarized light. This
plate can be rotated around the optical axis, and the images corresponding
to the two beams are registered for four angles: $0$, $22.5$, $45,$ and
$67.5$ degrees, allowing us to recover both the degree and direction of the
linear polarization at each point of the observed object. The narrow band (NB) filter
OIII+50 was introduced into the beam to select the Ly$\alpha$ emission, because
its central wavelength is perfectly suited to the redshift of the target.
The quasar has $z_{qso}=3.1092\pm 0.0014$, based on the H$_\beta$ and [O III]
emission line regions, while the Ly$\alpha$ emission line is
at $z_{Ly\alpha}=3.113\pm 0.001$ \citep{HPKZ09}; this corresponds to
a velocity difference of $277$~km\,s$^{-1}$. The redshifted nebular emission
corresponds to a wavelength $\lambda_0=5000.05$\,\AA, almost right on the
mean wavelength ($5001$\,\AA)
of the OIII+50 ESO filter. With $FWHM=57$\,\AA, this filter is wider than
the emission Ly$\alpha$ line ($FWHM\sim 8$\,\AA, see Fig.~1 of \cite{HPKZ09}),
so its blue and red wings clearly include part of the quasar
flux, even though it is strongly attenuated by the saturated Ly$\alpha$
absorption of the PDLA.
This raises the question of whether the quasar continuum may blur the polarization
signal; we address this question in Section \ref{sec:polar}.

A dithering pattern was adopted in the east-west
direction, which was parallel to the $x$-axis of the CCD detector and to the
long axis of the mask strips. Dithering in the other direction would have
led to a possible loss of information, since the width of the strips are
not much larger than the possible size of the Ly$\alpha$ blob. We used the
blue-sensitive EEV detector, because it is slightly more efficient
at $5001$\,\AA\ than the MIT detector; this is why we observed in visitor
mode.

In total, $100$ exposures (i.e., $25$ per HWP angle) were secured in dark time
(no Moon) and photometric conditions through the NB filter in the
IPOL mode, each with an exposure time of $500$\,seconds. The $1-\sigma$ noise in the background of each exposure is about $6$~ADU, clearly dominating the read-out noise (RON), which amounts to $2.2$~ADU. The airmass varied
between $1.30$ and $1.50$ for $83$ of them, and between $1.50$ and $1.85$
for the $17$ remaining ones. The seeing measured on the frames varied from $2.5$ pixels or $0.63$\arcsec to $4.2$ pixels or $1.05$\arcsec. On the combined intensity image, we measure a seeing of $0.76$\arcsec.

We also devoted about half an hour to secure eight exposures through
the wide-band v$_\mathrm{HIGH}$ filter, each with an exposure time of $200$\,s,
totaling $26$\,m,\;40\,s. The airmass varied between $1.34$ and $1.39$, and
the seeing measured on the combined image is $0.7$\arcsec.
Dithering was made in both the $x$ and $y$ directions. The deep image resulting
from the coaddition of all these frames is important for distinguishing sites
with continuum emission from those with Ly$\alpha$ emission.

\subsection{Reduction}
The reduction was made first using the FORS2 ESO pipeline (version 4.12.8)
in the \verb+gasgano+ (version 2.4.5) environment to make the master bias
and the master sky flat-field NB frames.
The sky flat fields were obtained without the Wollaston prism, HWP, and strip
mask, because including the polarization optics would be impractical, and is not required in view of the observation strategy,
as explained by \citet[in the Supplementary information section]{HSS11}.

Only the data registered by the
CHIP1 (Norma III) CCD was reduced, 
because
the quasar image always lies on that chip. Regarding the broad-band v$_\mathrm{HIGH}$ data,
we also focused on the CHIP1 frames, because half the CHIP2 (Marlene) images
were corrupted for some unknown reason. We used the \verb+fors_img_science+
recipe of the ESO FORS2 pipeline to reduce the v$_\mathrm{HIGH}$ CHIP1 images.
For the polarized NB images, the pipeline does not apply, so we had to use the
\verb+iraf+ command \verb+imarith+ to subtract the bias level and the master
bias, and divide by the master sky flat field.
For the NB images, we subtracted the sky in a preliminary way using the
\verb+sextractor+ code with a smoothing on 200 pixels. This subtraction was only approximate because slight but systematic and significant differences in the sky level (up to $\sim 1$~ADU) occur between the ordinary and extraordinary beams. Then, we
corrected the intensities to unity airmass, using the average Paranal extinction
coefficients listed in \citet[Table 3]{PMO11} ; we further scaled all images to the same flux level using the bright star closest to the QSO (1\arcmin33\arcsec from it) and close to the middle of CHIP1. To align the images, we applied
the \verb+alipy+ code\footnote{http://obswww.unige.ch/~tewes/alipy/apidoc/alipy.html}
written by Malte Tewes.
The \verb+alipy+ code applies an affine transform to the images.
We note that scaling amounted to less than a pixel at most.
We isolated the ten strips
corresponding to the 
ordinary and extraordinary beams. For each observing sequence and each
strip, we refined the sky 
subtraction for the eight subframes corresponding to the four angles and the two beams
using \verb+sextractor+ again.

Finally, the \verb+iraf+ command \verb+imcombine+ was applied to combine the aligned
images, using the \verb+comb=average+ and \verb+reject=minmax+ 
options. The combination was made separately for each angle of the 
half-wave plate.
The same procedure was applied to the intensity v$_\mathrm{HIGH}$ images,
resulting in a full image instead of five $22$\arcsec\ wide strip\footnote{Actually, the useable width of each strip is closer to $20$\arcsec than to $22$\arcsec.}.
The result is shown in Fig.~\ref{fig:morpho}, where the NB intensity images were obtained by further combining the eight subimages related to
the four angles and two beams.

\subsection{Flux calibration}
\subsubsection{Magnitude of the quasar through the $V_\mathrm{HIGH}$ filter}

We first verified that the quasar flux measured on our $V_\mathrm{HIGH}$ images fits the magnitude $V=19.57$ well \citep{VV10}. The magnitude obeys the relation
\begin{equation}
V_\mathrm{HIGH}=-2.5\,\log\left(\frac{N}{t_\mathrm{exp}}\right)+m_0
-K(V_\mathrm{HIGH})\cdot airmass
,\end{equation}
where $N$ is the electron count, $t_\mathrm{exp}$ the exposure time,
$m_0$ the instrumental zero point, and $K(V_\mathrm{HIGH})$ the extinction
coefficient.
On the combined (averaged) $V_\mathrm{HIGH}$ image, we measure
$N=534681$\,e$^-$ for $t_\mathrm{exp}=200$\,s, and the zero point is $m_0=28.19$
in February-March, 2014 for zero airmass according to the ESO QC1
database\footnote{archive.eso.org/qc1/qc1\_cgi}. We adopted the average extinction
coefficient $K(v_\mathrm{HIGH})=0.128$ \citep{PMO11} and obtain $airmass=1.37$.
With these numbers, we obtain $V_\mathrm{HIGH}=19.45$, brighter than
the catalog value. Actually, the zero point certainly refers to the
standard MIT detector rather than the EEV one we used, but at the mean
wavelength of the $V_\mathrm{HIGH}$ filter ($\lambda_0=5550$\,\AA), the quantum
efficiency of the EEV detector is only about $5$\% higher than that of the
standard MIT one \citep[Fig. 2.6]{BDK13}. One would then expect the
corresponding zero point to be about 0.05 magnitudes deeper for the EEV detector,
changing our magnitude to $19.50$. This is about $0.07$ mag brighter than that listed in Véron's catalogue.

\subsubsection{Absolute flux through the NB filter}
To calibrate our images in flux, we made use of the unpolarized spectrophotometric standard
EG274$\equiv$WD 1620-391 \citep{HWS92} that was observed on March 4, 2014 at 05:46 UT
through polarization optics under photometric conditions. It was observed four times by strongly variable
seeing, each time
through all four angles. Retaining the images with the best seeing\footnote{Although the total flux 
we considered does not depend on the seeing in principle, it does in practice because the star is
not centered in the middle of the $22$\arcsec strip, but it lies as close as $4\arcsec$ from the strip edge, so some flux is lost if seeing
is bad.}, we see that an
$68\,638$~ADU\,s$^{-1}$ count rate (obtained by integrating the signal in an aperture of $17$
pixels radius\footnote{This radius is chosen on the basis of a curve of growth (aperture photometry); limiting the radius to this value implies
a flux loss of no more than two per mil, while limiting it to, for example, 12 pixels (3”) would imply a loss of about 1.8 \%.},
i.e., $4.27$\arcsec, and correcting for a $k=0.156$ atmospheric extinction, with $airmass=1.56$)
corresponds to an average flux of
\begin{equation*}
\frac{\int F(\lambda)\,T (\lambda)\,d\lambda}{\int T(\lambda)\,d\lambda}
=1.92\times 10^{-13}\,\mathrm{erg\,s}^{-1}\,\mathrm{cm}^{-2}\,\mathrm{\AA}^{-1,}
\end{equation*}
computed through the filter transmission curve $T(\lambda)$ kindly provided by ESO.
Multiplying this flux by the filter ``equivalent width'' defined as
$W_{NB}=\int_{4800}^{5200} T(\lambda)\,d\lambda=41.7$\,\AA, one obtains a total flux of $8.0\times 10^{-12}$\,erg\,s$^{-1}$\,cm$^{-2}$. Here, the
$68\,638$~ADU\,s$^{-1}$
count rate is measured in {\sl \emph{one}} beam (ordinary or extraordinary) and corrected to
zero airmass.
When only one beam is considered (and assuming unpolarized light),
the conversion factor is then
\begin{equation}
f_\mathrm{conv}=1.17\times 10^{-16}\,\mathrm{erg}\,\mathrm{cm}^{-2}\,\mathrm{ADU}^{-1}
\label{eq:conv}
,\end{equation}
with an uncertainty that we estimate to roughly $10$\%.

\subsection{Polarization calibration}
\label{subsec:polcal}
We computed the polarization fraction and angle (according to the formulae
\ref{eq:polar_measured}, \ref{eq:F_formula}, and \ref{eq:QU_formulae}; see below)
of the polarized star Vela1~95$\equiv$Ve$6-23\equiv$GSC$08169-004127$, observed on March 3, 2014
through the NB OIII filter. The star image fell on the same position
on the CCD as that of the J1240 QSO target. We obtained
\begin{equation}
P=8.13\pm 0.13~\%~~~;~~~~\theta=172.9\degr\pm 0.5\degr
\label{eq:Vela1_23}
,\end{equation}
where the angle $\theta$ has been corrected for the instrumental effect\footnote{See
VLT-MAN-ESO-13100-1543, issue 96.0, p.~40, Fig.~4.1.}
$\varepsilon_\mathrm{F}\simeq 3.5\degr$
according to the relation $\theta=\theta_\mathrm{obs}-\varepsilon_\mathrm{F}$, and
where the errors are estimated through simplified formulae \citep[][Eqs. 4, 9]{FBM07}.
This is consistent with measurements made with the FORS1 instrument in the IPOL mode
through the broad-band $V$ filter by
\citet[][$P=8.26\%\pm 0.05$\%, $\theta=171.61\degr\pm 0.21$\degr]{FBM07}.
More importantly, this is also consistent with observations made with the
FORS2 instrument in the PMOS mode, which allows one to determine $P$ and $\theta$
as a function of wavelength \citep{CPC16}. Indeed, both $P$ and $\theta$ depend
on wavelength, and the mean wavelength of the $V$ filter is $\sim 5500$\,\AA,\ while
that of the OIII+50 ESO filter is $5001$\,\AA. At the latter wavelength,
$P=8.0\%\pm 0.1$\% and $\theta=172.8\degr\pm 0.3$\degr\ \citep[][Figs. 3, 6]{CPC16}.
Thus, our measurements recover $P$ to within the $1-\sigma$ error bars.

To assess the lack of substantial instrumental polarization, we also considered the
non-polarized star WD1620-391, which has been measured through
both the OIII+50 NB filter (on March 2, 2014) and the ESO b$_{HIGH}$ wide-band
filter (on March 4, 2014). Through the OIII+50 filter, we obtained $P=0.11\%\pm 0.15$\% and
$0.18\%\pm 0.15$\% for the two measurements with good seeing, while we obtained
$P=0.07\%\pm 0.07$\% through the v$_{HIGH}$ filter. Likewise, the unpolarized
star WD1615-154 observed through the 
OIII+50 filter on March 5, 2014 gives $P=0.45\%\pm 0.14$\% and
$0.28\%\pm 0.14$\% for two measurements with excellent seeing.
All this is compatible with zero polarization.

\subsection{Polarized sources in the field}
We applied the \verb+Sextractor+ code to the average images
corresponding to each of the two beams, four angles, and five 22\arcsec\
bands registered on CHIP1. The polarization fraction and angle were
computed (see formulae in Section~\ref{sec:polar}) for each source detected in all average images, from the fluxes
determined by \verb+Sextractor+. The errors on the polarization fraction
$P$ were computed from the flux errors and corrected as explained in
Section~\ref{sec:polar}. The polarization angles were corrected for the
instrumental effect mentioned in Section~\ref{subsec:polcal}.
The results are listed in Table~\ref{tab:pol_all} for all objects,
and shown only for objects with $P > 5\,\sigma(P)$ as red
segments in Fig.~\ref{fig:morpho}.

The polarization of some objects is very small and probably spurious,
because the error estimate does not include systematic effects: such is the case of the very bright star to the west of the QSO,
close to the right edge of the frame. On the other hand, several
sources have large $P$ values. Their spatial distribution in the field
does not suggest any connection with the QSO, and the orientations of
their polarization vectors seem random. The most polarized object in
the field, however, is close to the QSO in the field, and its 
polarization vector is tangential relative to the direction of the 
QSO. Thus, one might speculate that, if this object was at the same
distance as the QSO from us, its polarization could be due to 
scattering of the the QSO Ly$_\alpha$ radiation by its hydrogen halo.
Measuring the redshift of this object would help settle the question.

\section{Morphology, luminosity, and surface brightness}
\label{sec:results1}

\subsection{Subtraction of the quasar signal in the NB images}
Part of the signal in the NB image is due to the quasar itself, because the filter
is not narrow enough to exclude the wings of the saturated Ly$\alpha$ absorption
of the PDLA. Therefore,
we had to subtract this contribution before analyzing the properties of the LAB.
Since there are four polarization angles, there are four combined NB images, each
containing two subimages corresponding to the ordinary and extraordinary beams;
this makes a total of eight subimages per polarization measurement sequence. In total, we obtained 25 exposures for each polarization measurement sequence. Co-adding those 25 exposures, the deepest possible intensity image reaches a $1-\sigma$ noise of $\sim 0.53\times 10^{-18}~\mathrm{erg}\,\mathrm{s}^{-1}\,\mathrm{cm}^{-2}\,\mathrm{arcsec}^{-2}$ in the sky. The mean seeing of our data is 0.76 \arcsec FWHM.

The quasar/LAB light decomposition is performed as follows. First, the PSF is reconstructed using the {\tt STARRED} PSF reconstruction and deconvolution Python package \citep{Michalewicz2023}. A subsampled (by a factor of 2) model of the PSF is generated for each exposure, and for each of the two beams and four position angles of the Wollaston plate by fitting the light profile of two bright stars in the field of view. {\tt STARRED} obtains the best-fit model of the PSF by first adjusting a Moffat profile to those stars and then applying a pixelated correction, regularized with a wavelet. 
Second, we use {\tt STARRED} to perform a two-channel deconvolution of the data in order to separate the quasar light from the extended emission. For each of the two beams and four position angles, we perform a joint deconvolution of all the 25 exposures using individually reconstructed PSFs in each image. {\tt STARRED} models the quasar as a point source and the LAB extended emission as a wavelet-regularized pixel grid. These two channels are jointly fit to the data. The amplitude of the point source and a spatially constant sky correction is free to vary between individual exposures, whereas the extended emission channel is kept constant across all exposures. The reconstruction is performed at twice the resolution of the original data, and we also fit a small sub-pixel shift to all exposure to align the images. 
Finally, we subtract our best-fit point-source model to each of the images and stack the 25 exposures before performing our polarization measurement. {\tt STARRED} also provides the photometry of the quasar for each beams and position angles. We measure a polarization for the quasar of $P=0.0052\pm 0.0053$, which is consistent with previous measurements of radio quiet quasars. Indeed, while radio galaxies do 
show important polarization fractions of up to $0.15-0.20$
\citep{CdV98,VFV99}, most radio quiet quasars (like ours) have 
$P<0.01$; according to \citet[Fig.~1]{HBP14}, $86$\% of the RQQs have 
$P<0.01$ and $97$\% have $P<0.02$.

\subsection{Morphology}
\begin{figure*}[ht!]
\centering
\includegraphics[width=17.6cm]{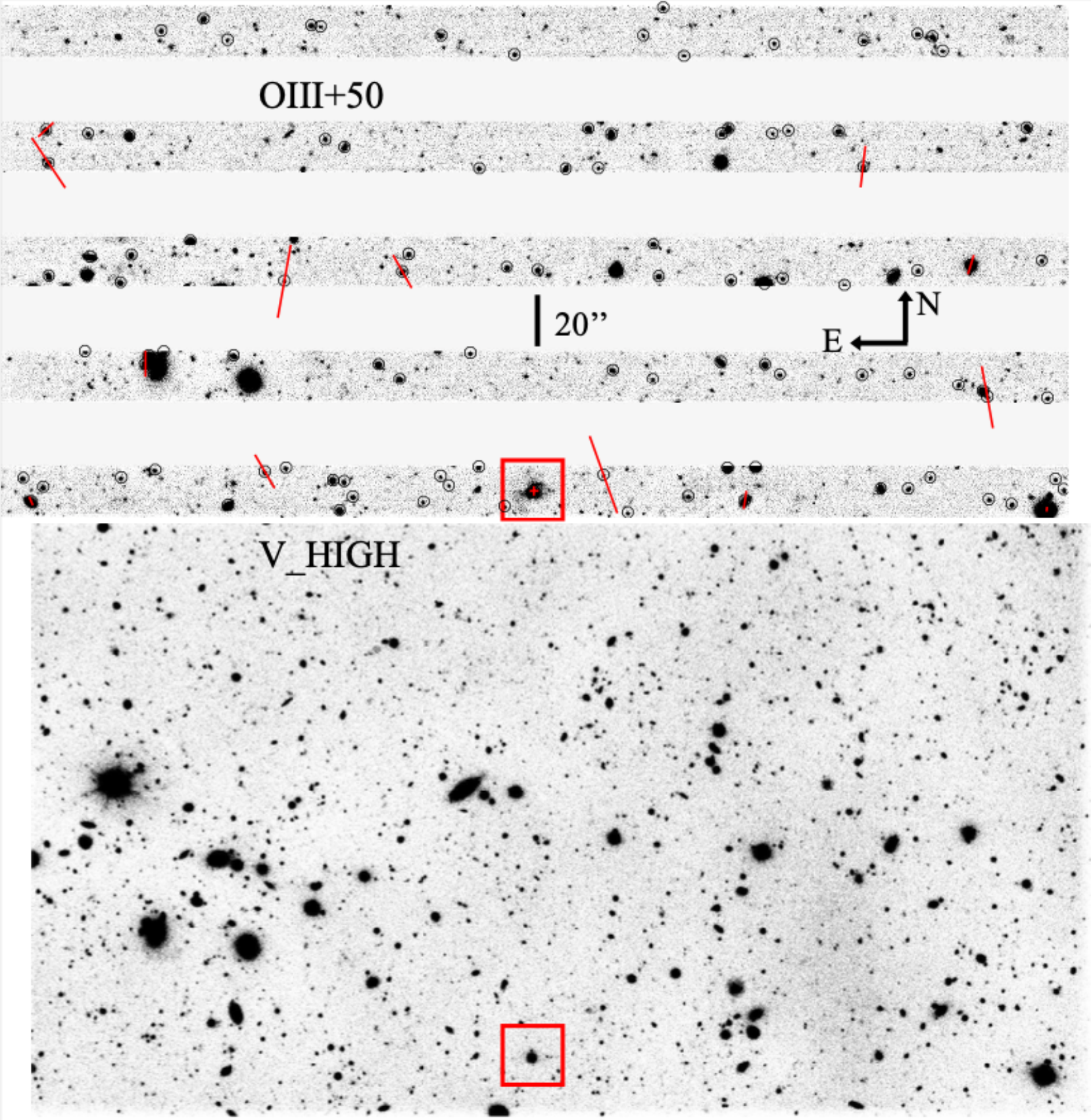}
\caption{Field covered by the CHIP1 detector, through the NB filter and through the wide-band filter.
{\bf Upper panel:} Total intensity image in redshifted Ly$\alpha$
band, obtained by coadding all images corresponding to all four polariser angles
and to both ordinary and extraordinary beams, as recorded on the CHIP1
detector. The total equivalent exposure time
is 13 hours, 53 minutes, and 20 seconds.
The five bands correspond to the positions of the ordinary beam; the bands where the extraordinary beam falls appear blank here, because the corresponding images have been aligned and stacked with those of the ordinary beam. This shows that only half of the field is accessible in
the polarization mode. The red square indicates and surrounds the LAB image, which includes the overexposed QSO near its center. The black circles designate objects detected at each angle and beam, and for which
the polarization was measured. The red segments are the polarization vectors for objects with a polarization fraction of $P > 5\,\sigma$.
{\bf Lower panel:} Total v$_\mathrm{HIGH}$ intensity image,
corresponding to a total exposure time of 26 minutes, 40 seconds. The same red square as in the upper panel identifies the QSO.
Both frames have the same scale.}
\label{fig:morpho}
\end{figure*}

 The average intensity image is shown in Fig.~\ref{fig:morpho}
and reveals the shape of the blob. Interestingly, this deep exposure shows
that, although the bright central part of the blob is no larger than $\sim 5$\arcsec\
as mentioned by \citet{HPKZ09}, fainter structures extend up to radii of
$\sim 8$\arcsec. Thus, the total extent of the blob reaches $\sim 16$\arcsec,
or $122$~kpc (taking into account that $1$\arcsec
corresponds to $7.615$\,kpc). With its complex and filamentary structure, this blob appears quite typical of those found around radio-quiet quasars using the MUSE integral field spectrograph \citep{BCL16}.

Several small and faint objects surrounding the blob are seen through both the
OIII+50 and $V_\mathrm{HIGH}$ filters and may be small galaxies.
The faint patch at $8$\arcsec\ to the SW of the quasar has no
counterpart on the v$_\mathrm{HIGH}$ image. It may be a hydrogen cloud
of the same kind as those detected by \citet{CLH12}.

\subsection{Luminosity of the blob}
Within a radius of $30$\,pixels or $7.53$\arcsec, the sum of the pixel values of
the nebular emission amounts to $24.82$\,e$^{-}$\,s$^{-1}$ outside the atmosphere, and as
a first approximation we consider it negligible beyond this radius.
Applying the conversion factor
$f_\mathrm{conv}$ (\ref{eq:conv}) and remembering that
$1~\mathrm{ADU}=2.24~\mathrm{e^{-1}}$, we obtain a total flux of
\begin{equation}
F(Ly\alpha)\simeq (1.30\pm 0.13)\times 10^{-15}\,\mathrm{erg}\,\mathrm{s}^{-1}\,\mathrm{cm}^{-2}.
\label{eq:totflux}
\end{equation}
The uncertainty was set at $\sim 10$\%, essentially
reflecting the uncertainty on the flux calibration.
Interestingly, the intensity
of the Ly$\alpha$ emission shows two maxima, one that occurs at the position of the quasar and another that is slightly offset by about $0.6$\arcsec\ to the east and $0.3$\arcsec\ to
the south of the quasar.
The positions of the quasar and of the secondary maximum
are summarized in Table~\ref{qso_lab}.
Integrating the surface brightness profile
fit to the data up to a $7.53$\arcsec\ radius (see below) resulted in the
same value as that given above (Formula \ref{eq:totflux}), within $11$\%. 

The luminosity distance is
$26572$\,Mpc \citep{W06}\footnote{http://www.astro.ucla.edu/$\sim$wright/CosmoCalc.html}.
The luminosity of the blob is thus
\begin{equation}
L(Ly\alpha)\simeq (1.09\pm 0.11)\times 10^{44}\,\mathrm{erg}\,\mathrm{s}^{-1},
\end{equation}
a value close to the median one for the blobs in the sample of \citet{BCL16}.
Among the $17$ blobs of this kind observed by these authors, seven are brighter than our object and ten are fainter\footnote{Figure~1 of \citet{HPKZ09} shows that the width of the Ly$\alpha$ emission line is much narrower than that of our NB filter, so that no light is lost in our luminosity estimate. Besides, Fig.~2 of \citet{BCL16} shows similarly narrow emission lines for all $17$ LABs surrounding RQQs.}.
Besides, only one of the $14$ LABs listed by \citet[Table 2]{MYH11}, SSAA22-Sb3-LAB1, is
brighter than that. We note that the two articles quoted here use the same
cosmological parameters as we do.

\begin{table*}
\caption{Characteristics of quasar SDSS~J$124020.91+145535.6$
and position of the secondary maximum of the LAB intensity.}
\begin{center}
\begin{tabular}{lllcc}
\hline\hline
           &\multicolumn{2}{c}{Coordinates J2000}& $B$ mag& $V$ mag  \\
           &$\alpha$ (h mn s)&$\delta$ ($\deg$ ' $\arcsec$)&   &     \\ \hline
Quasar     &$12$ $40$ $20.914$&$+14$ $55$ $35.73$& $20.15$ & $19.57$ \\
LAB 2ndary max. &$12$ $40$ $20.942\pm0.004$&$+14$ $55$ $35.4\pm0.06$&  &  \\  \hline
\end{tabular}
\end{center}
\label{qso_lab}
\end{table*}

\subsection{Surface brightness}
\label{surf_bright}
\begin{figure}[ht!]
\centering
\includegraphics[width=8.8cm]{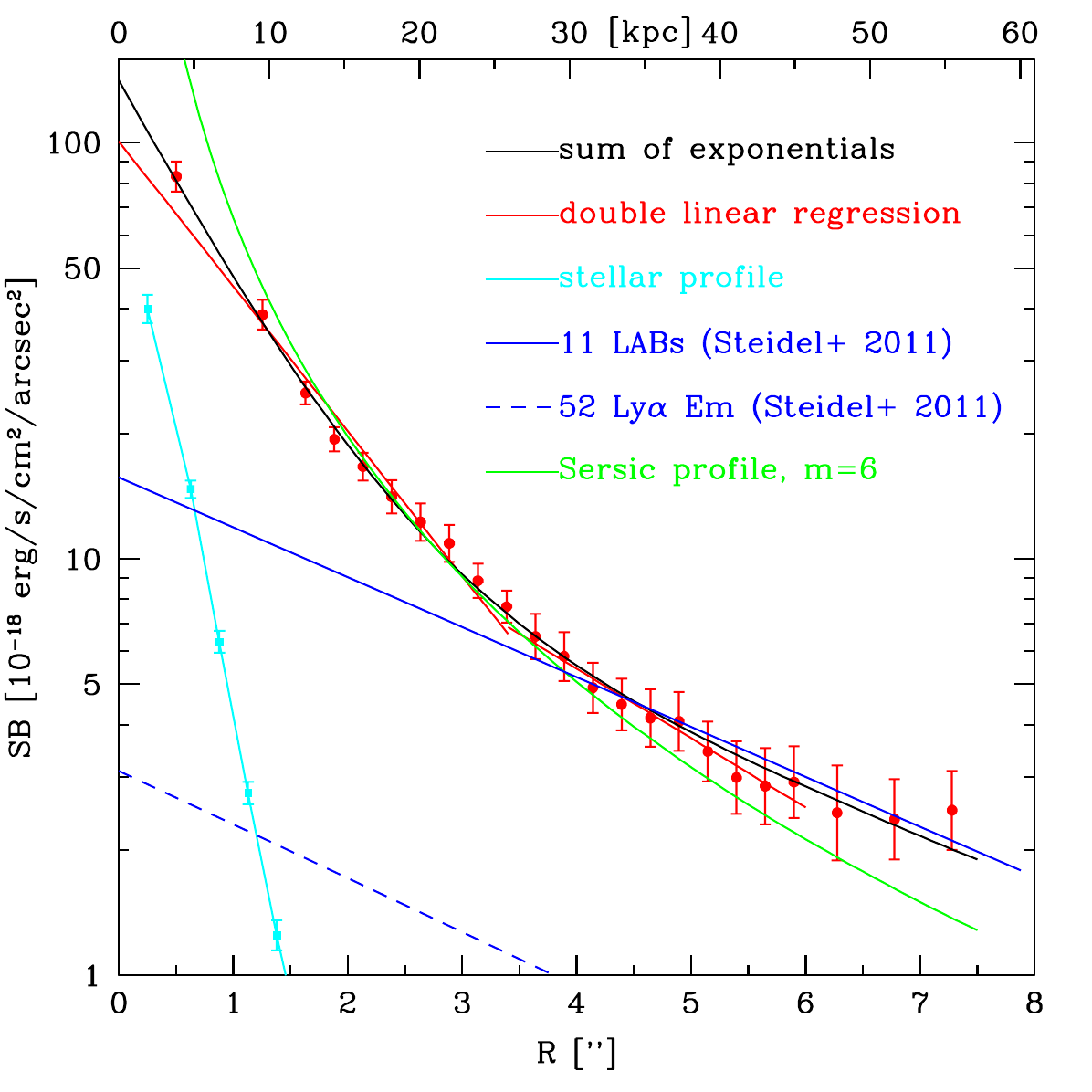}
\caption{Average surface brightness (SB) as function of radius. Each radius value is
the arithmetic mean of the two bounding radii of the corresponding annulus (e.g.,
$R=2\,\mathrm{pixels}\simeq 0.5\arcsec$ for the central $R<4$\,pixel disk and
$R=5$\,pixels for the $4\leq R<6$\,pixel annulus).
The error bars represent the rms scatter of the pixel values in the corresponding
annulus; it includes not only the shot noise, but also the cosmic SB variation
within the annulus. The two red regression lines
are those given by Eqs. \ref{eq:sb_rad1} and \ref{eq:sb_rad2}.
The black curve is the fit of a sum of two exponentials (Eq.~\ref{eq:sb_rad_addexp}), while the green one is a S\'ersic profile with index $m=6$;
the cyan curve is a
stellar profile (see text). The solid blue  line is the fit to the average SB profile of the 11 LABs
considered by \citet{SBS11}, while the broken blue line is the same for the 52 galaxies
with Ly$\alpha$ in emission \citep{SBS11}.
}
\label{fig:sb}
\end{figure}

We obtain the surface brightness (SB) profile displayed
in Fig.~\ref{fig:sb} by adding the pixel values in concentric annuli with increasing
radii (the first ``annulus'' is a disk with radius $r=4$\,pixels, while the
annuli proper are two pixels thick between $r=4$\,pixels and $r=6$\,pixels, one pixel thick between
$r=6$\,pixels and $r=24$\,pixels, and two pixels thick between $r=24$\,pixels and $r=30$\,pixels).
We adopted the quasar as the center of the photometric apertures.

For comparison, in Fig.~\ref{fig:sb} we also show the SB profile of a star (equivalent to a PSF) in the same
field of view and in the same strip as the QSO in the combined NB image. The star used
lies at $\alpha(J2000)=12:40:26.246$ and $\delta(J2000)=+14:55:27.7$, and its SB
within $2$\,pixels ($r<0.5$\arcsec) was scaled to coincide with that of the LAB
at same radius. The contrast between the two curves confirms the
extended nature of the object.

 The SB trend of the blob shows two linear
parts in this logarithmic plot, the first within $3.4$\arcsec\ ($26$\,kpc)
with a steep slope and the second between $3.4$ and $6$\arcsec\ ($26$ and $46$\,kpc)
with a shallower slope\footnote{We note that this SB profile refers only to the Ly$\alpha$
flux, as measured through the NB filter, not to the continuum.}. Assuming a halo mass between $10^{12}$ and $10^{13}~M_\sun$ at $z=3.1$, the virial radius would correspond to $\sim 80-170$~kpc, so we are probing a region that is between $1/3$ and $2/3$ of the virial radius, for the most extended points.
The error bars correspond to the rms dispersion of the pixel
values within each annulus. Thus, they include not only the photon noise, but
also the cosmic SB variation within the respective annuli, so they are larger than
the error due to the shot noise alone. Although the annuli are disjointed, their SB
and error bars are correlated because of the seeing.

The regression lines were fit using weighted least squares (the weights being
the inverse of the variances, but neglecting the correlations) and obey the following laws:
\begin{eqnarray}
\label{eq:sb_rad1}
\log(SB)&=&-15.997\pm0.047-(0.348\pm0.021)\cdot R\,[\arcsec] ,\\ \nonumber
&& ~~~\mathrm{for}~~ R < 3.4\arcsec \\
\label{eq:sb_rad2}
\log(SB)&=&-16.596\pm0.048-(0.167\pm0.010)\cdot R\,[\arcsec] ,\\ \nonumber
&& ~~~\mathrm{for}~~ 3.4\arcsec < R < 6\arcsec
\end{eqnarray}
with SB expressed in
$\mathrm{erg}\,\mathrm{s}^{-1}\,\mathrm{cm}^{-2}\,\mathrm{arcsec}^{-2}$.
The rms scatter of the residuals is $0.049$ and $0.020$\,dex for the inner
and outer parts of the LAB, respectively.
The SB beyond $R=6$\arcsec\ lies increasingly above the linear trend. Even though
the last three points correspond to a very low SB, they probably represent
a real flux, as suggested by the large, very faint structures still visible
up to $\sim 8$\arcsec\ from the quasar in Fig.~\ref{fig:morpho}.
Integrating Eqs.~\ref{eq:sb_rad1} and \ref{eq:sb_rad2}
up to $R=6$\arcsec\ results in a total flux of $1.06\times 10^{-15}\,\mathrm{erg}\,\mathrm{s}^{-1}\,\mathrm{cm}^{-2}$,
about $18$\% smaller than that obtained
(Eq.~\ref{eq:totflux}) from direct summation of the signal, as expected because
some signal remains beyond $R=6$\arcsec.

Taking into account that $1$\arcsec\ corresponds to $7.615$\,kpc, the SB can also
be written as follows:
\begin{eqnarray}
SB&=&(10.08\pm1-09)\cdot 10^{-17}\,\exp\left(-\frac{R\,\mathrm{[kpc]}}{9.51\pm 0.58}\right) ,\\ \nonumber
&& ~~~\mathrm{for}~~ R < 26\,\mathrm{kpc} \\
SB&=&(2.53\pm0.28)\cdot 10^{-17}\,\exp\left(-\frac{R\,\mathrm{[kpc]}}{19.8\pm 1.2}\right) ,\\ \nonumber
&& ~~~\mathrm{for}~~ 26\,\mathrm{kpc} < R < 46\,\mathrm{kpc}
\label{eq:sb_rad_kpc}
\end{eqnarray}
with $R$ expressed in kpc and $SB$ still expressed in
$\mathrm{erg}\,\mathrm{s}^{-1}\,\mathrm{cm}^{-2}\,\mathrm{arcsec}^{-2}$.

Although the double linear regression just described fits the data very well, it
suffers from the drawback of an {\sl \emph{a priori}} definition of the regression limit in radius
($3.4$\arcsec) that is defined by eye in a rather subjective way. A more objective
fit was performed to the sum of two exponential functions (actually
to the decimal logarithm of it), giving:
\begin{equation}
\begin{split}
SB=(129.0\pm8.0)\cdot 10^{-18}\cdot\exp\left(-\frac{R\,[\arcsec]}{0.818\pm0.051}\right) \\ 
+(12.7\pm 2.9)\cdot 10^{-18}\cdot\exp\left(-\frac{R\,[\arcsec]}{3.93\pm0.61}\right)
\,\mathrm{erg\,s^{-1}\,cm^{-2}\,arcsec^{-2}}
\end{split}
\label{eq:sb_rad_addexp}
,\end{equation}
or, translating arcseconds into kiloparsecs,
\begin{equation}
\begin{split}
SB=(129.0\pm8.0)\cdot 10^{-18}\cdot\exp\left(-\frac{R\,[kpc]}{6.23\pm0.39}\right)\\
 +(12.7\pm 2.9)\cdot 10^{-18}\cdot\exp\left(-\frac{R\,[kpc]}{29.9\pm4.7}\right)
 \,\mathrm{erg\,s^{-1}\,cm^{-2}\,arcsec^{-2}}
 \end{split}
\label{eq:sb_rad_addexp_kpc}
,\end{equation}
with an rms scatter of the logarithmic residuals of $0.031$\,dex. The relatively large errors on the fit
parameters stem from the inclusion of the last three points, which have among the largest
error bars, and from correlations between some parameters (in particular between the
two amplitudes, and between the amplitude and scale of the second exponential term).

\begin{figure}[ht!]
\centering
\includegraphics[width=8.8cm]{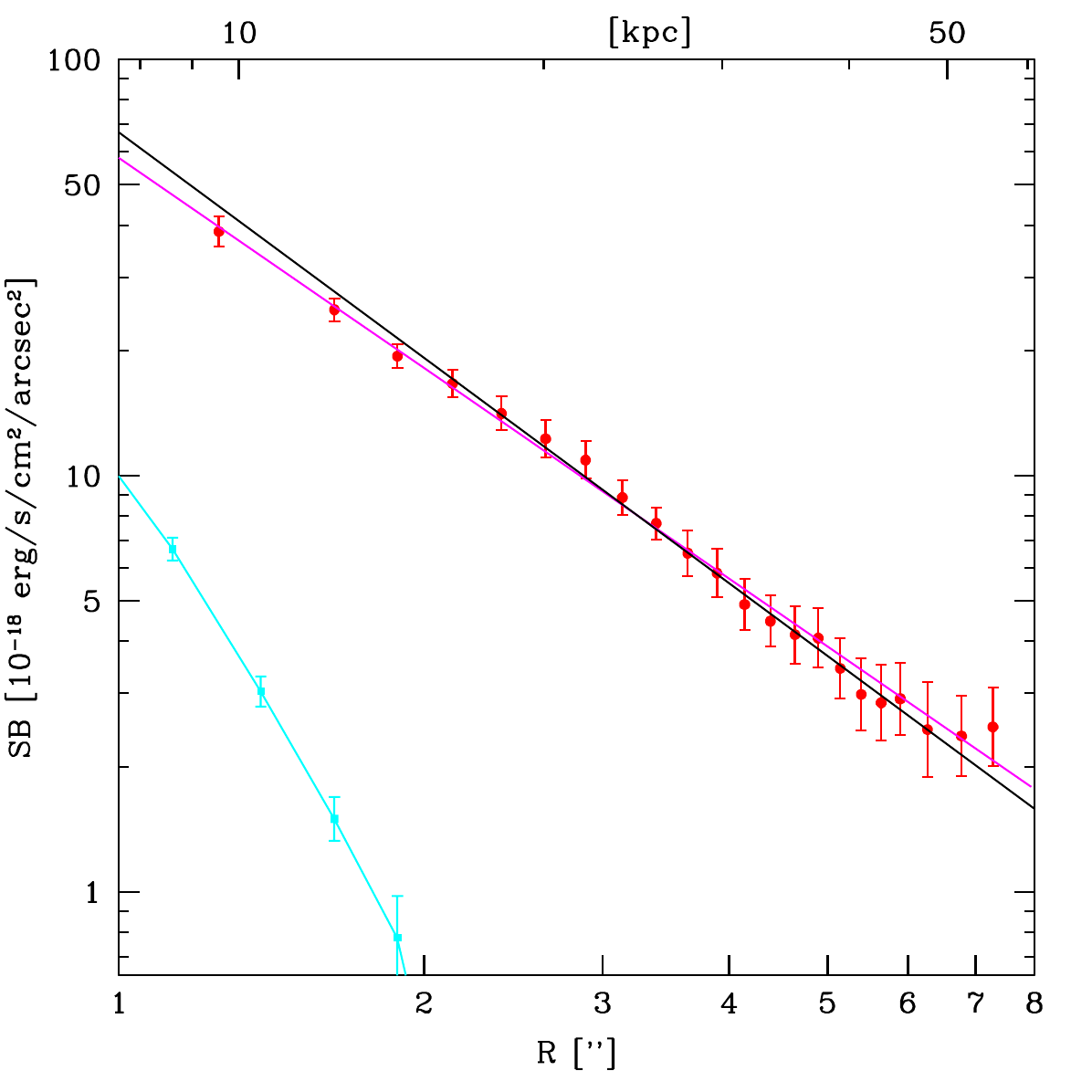}
\caption{Same as Fig.~\ref{fig:sb}, but with a logarithmic scale on both axes, and
starting only from $R=1$\arcsec. The regression line shown in magenta is computed from
the red dots displayed here, discarding the dot at $R\simeq 4$~kpc. The
black line shows the power law with an index $-1.8$, corresponding to the average
SB profiles of the blobs studied by \citet{BCL16}.
The stellar profile is shown in cyan.}
\label{fig:sbll}
\end{figure}

\begin{figure*}[t!]
\centering
\includegraphics[width=18cm]{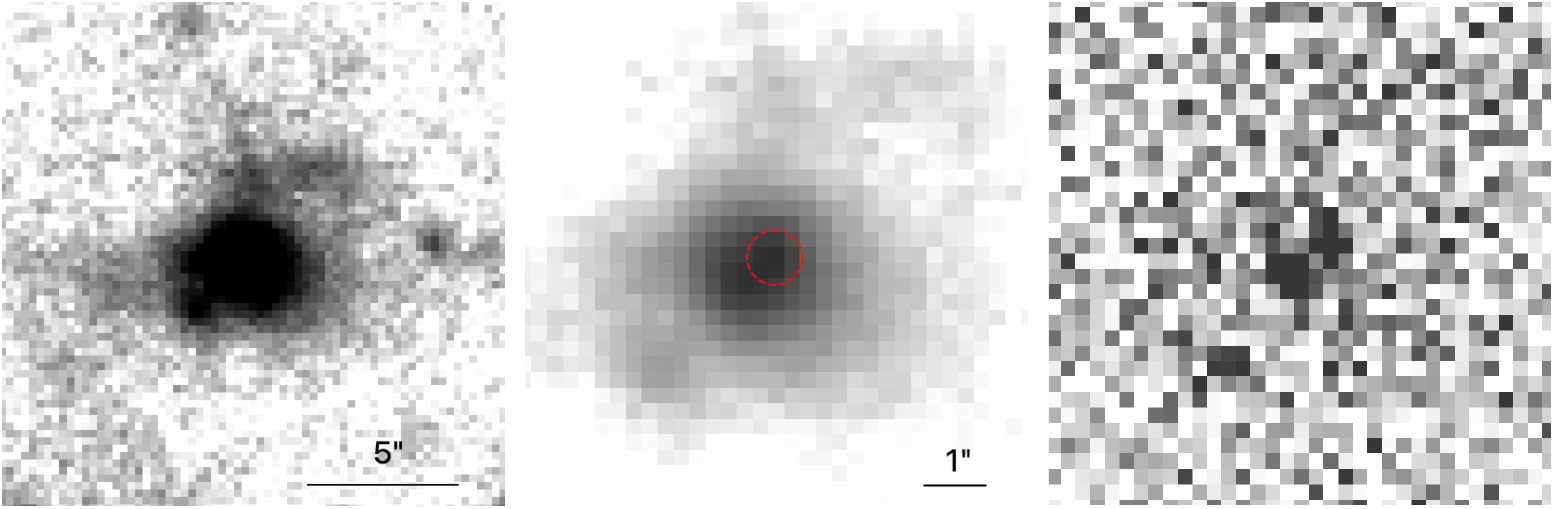}
\caption{Central structure of LAB, with north up and east left. On the left is the extended low-surface-brightness structures around the quasar. The middle panel shows a zoomed-in view of the central parts after subtraction of the point-source component of the image following the deconvolution of all frames with STARRED. The image is a stack of all quasar-subtracted images (see text) and the position of the point-source component is shown as a red circle. This image is at the spatial resolution of the original data. We note the double structure of the flux distribution. The north-west maximum corresponds to the position of the quasar that has been subtracted, while the SE maximum
has no direct connection with the quasar. The right panel shows the stacked residuals with cut levels spanning -3$\sigma$ to +3$\sigma$. We note that while a structure is seen in the center, it is quantitatively negligible.
}
\label{fig:blob_centre}
\end{figure*}

\subsubsection{Comparison with other LABs and Ly$\alpha$ haloes of high-z galaxies}
The SB profile of the inner part of our blob is 
thus much steeper than that of the $11$ giant Ly$\alpha$ blobs considered by \citet[Table\,2, column 6]{SBS11},
that have an e-folding radius\footnote{that is, the radius where the $SB$ is a factor $e$ lower than the
central $SB$ value.} of $27.6$\,kpc (i.e., more than three times larger).
It is also steeper than the average diffuse Ly$\alpha-$ emitting halo of the $52$ high-redshift
star-forming galaxies (with net Ly$\alpha$ emission) studied by \citet{SBS11}, which have a typical radius
of $25.6$\,kpc, quite close to that of the giant blobs.
On the other hand, the outer part of our blob has an SB profile closer to that reported by \citet{SBS11}, though the uncertainty is large. 
Interestingly, the SB of the diffuse Ly$\alpha$ halos of the galaxies studied by
\citet[Figs.~5-8,]{SBS11} also show 
a steep linear part (on a log scale) in the inner region and a less steep linear
relation in the outer region, the break occurring between 25 and 30~kpc, which is only
slightly more than in our object. This is intriguing, because Steidel's galaxies
generally do not currently host any active nucleus, in contrast to SDSS~J124020.91+145535.6.
On the other hand, the surface brightness of our object is typically an order of
magnitude higher than that of Steidel's objects. This can be interpreted in two
ways. First, while Steidel's objects are representative of
$L\simeq L_\star$ Lyman break galaxies (LBG), our object is likely representative
of more luminous galaxies, since it hosts a quasar \citep{HCC84,SHB86,PMM01,HCT02,
HKO07,FBK14}. Second, the quasar that is present in our object but not in
Steidel's probably enhances the nebular Ly$\alpha$ emission by photoionizing
hydrogen \citep{OND13, CMBG08}.

The LAB1 object, which seems to be included in Steidel's sample of $11$ giant LABs, was studied
in detail by \citet{HSS11}. It has an e-folding radius of about $20$\,kpc beyond $\sim 2$\arcsec, close to the average of the $11$ LABs. As the latter, it does not show the double exponential behavior that is
present in both our object and Steidel's stacked galaxies.

\citet{WBB16} observed $26$ Ly$\alpha$-emitting galaxies at large redshift
($3<z<6$) with the MUSE spectrograph and detected a Ly$\alpha$ halo around most
of them. The SBs of these haloes follow an exponential law, but with a typical
characteristic length of only a few kiloparsecs. \citet{LBW17} confirmed this result on a much larger sample of star-forming galaxies, finding an average scale length of $4.5$~kpc. They are much less extended than our object.

These comparisons remain limited by the unknown contribution of a possible UV
continuum, especially near the center of the object. Nevertheless, the SDSS spectrum
displayed in Fig.~1 of \citet{HPKZ09} corresponding to $R<1.5$\arcsec suggests that
such a continuum is small, if any.

\subsubsection{Comparison with other quasar Ly$\alpha$ haloes}
Interestingly, discarding the central SB and plotting $\log(SB)$ against $\log R$
leads to an almost linear relation with a slope $-1.68\pm 0.03$ (Fig.~\ref{fig:sbll}),
in fair agreement with the average slope of $-1.8$ found by \citet{BCL16} for
$17$ blobs surrounding bright radio quiet quasars. \citet{BCL16} consider only the
range $R>10$~kpc, just as in our Fig.~\ref{fig:sbll}, and find that a power law
better fits their data than a single exponential.

The reason why a power law appears more satisfactory to \citet{BCL16} may be due
to their blobs obeying a double exponential law, as is the case of our object.
It would be interesting to reanalyze their data and look how far a double exponential
is able to fit them.

\subsubsection{Other possible fits to the SB radial profile}
We tried to adjust a S\'ersic profile,
\begin{equation}
\mathrm{SB}(R)=I_\mathrm{e}\,\exp\left(-b_\mathrm{m}\left[\left(\frac{R}{R_\mathrm{e}}\right)^{1/m}-1\right]\right)
,\end{equation}
where $R_\mathrm{e}$ is the half-luminosity radius and $I_\mathrm{e}$ is the surface brightness at that radius. We used Eq.~\ref{eq:sb_rad_addexp} to determine these two parameters: $R_\mathrm{e}=2.73$\arcsec, $I_\mathrm{e}=10.9\cdot 10^{-18}\mathrm{erg}\,\mathrm{s}^{-1}\,\mathrm{cm}^{-2}\,\mathrm{arcsec}^{-2}$. We also used the relation $b_\mathrm{m}=2m-0.324,$
which is a very good approximation to the real one \citep{BT08,CB99}. For all Sersic indices
$m=2-6$, the fit is acceptable for $R\geq 1.7$\arcsec, but the central intensity of the
profile diverges to values far above the measured one. The profile for $m=6$ is shown on Fig.~\ref{fig:sb}. This type of profile
clearly does not fulfill the observational constraints.

\subsection{Central part: The PDLA against the diffuse Ly$\alpha$ background}
The LAB structure is shown in more detail in 
Fig.~\ref{fig:blob_centre}.
The left panel shows the faint extended features, while the central
panel shows the inner part of the LAB. The red circle shows the position
of the quasar that has been subtracted through the deconvolution
process. The right panel shows the stacked residuals left by the
deconvolution, after subtraction of both the extended and point-source
components.

The middle panel of Fig.~\ref{fig:blob_centre} reveals that the central 
part of the LAB does not present a single maximum, but rather a bimodal
structure with one maximum coinciding with the quasar position and
another maximum to the south-east of it. These two positions are
summarized in Table~\ref{qso_lab}. The reason for this asymmetry is
unclear. \citet{HPKZ09} suggested that we might be seeing the PDLA silhouetting against the bright background.
These authors argued that star
formation in the PDLA cannot account for the extended Ly$\alpha$ emission in \ion{H}{ii}
regions, unless it is unrealistically large. They also argue that fluorescent
recombination radiation taking place in the PDLA ISM and powered by the quasar
cannot be invoked either, because the optically thick gas giving rise to the Ly$\alpha$
emission shields the same radiation from the observer's view.
Thus, one may wonder whether the asymmetry of the flux 
distribution is related with the PDLA, that would block not only the
Ly$\alpha$ flux of the quasar but also part of that of the LAB.
Our data has not enough spatial resolution to clarify this point; it
is compatible as well with a mere asymmetry of matter
distribution within the LAB.
The adaptive optics and narrow-field mode of the MUSE instrument is perfectly
suited to explore the central region of this LAB, and it might be able
to shed light on this issue.

\section{Polarization}
\label{sec:polar}
\begin{figure*}[ht!]
\centering
\includegraphics[width=18cm]{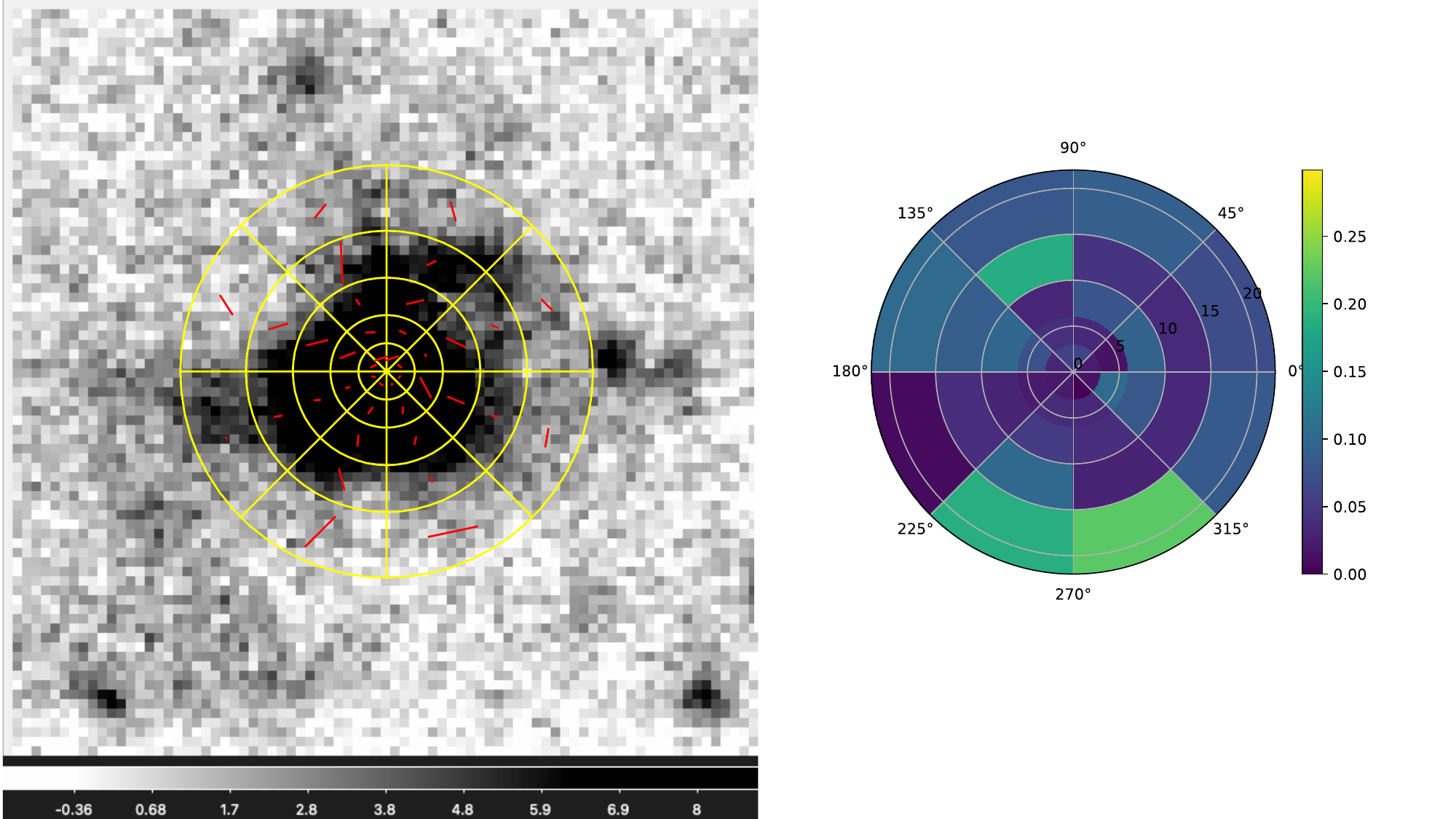}
\caption{Polarization signal in sectors of successive annuli of increasing
radius around the quasar. {\bf Left:}  Definition of the sectors on the average LAB image. The first ``annulus'' is actually a disk with a three-pixel
($0.75$\arcsec) radius, while the other annuli have an outer radius of $1.5$, $2.5$, $3.75,$
and $5.5$\arcsec,\ respectively. The red segments indicate the polarization fraction
through their length and the polarization angle through their orientation.
The background image is a median stack of the $200$ individual
images corresponding to the two beams, four angles, and $25$ exposures. The corresponding intensity bar is in electron units, representing the
average flux received during $500$ seconds and corrected for atmospheric
extinction.
{\bf Right:}  Map of polarization fraction. The latter was corrected statistically
using Eq. (A3) of \citet{WK74}, but spurious polarization still appears in some sectors
with low signal (e.g., the outermost one between $270$ and $315^\circ$).}
\label{fig:P_sector}
\end{figure*}
\begin{figure}[ht!]
\centering
\includegraphics[width=8.8cm]{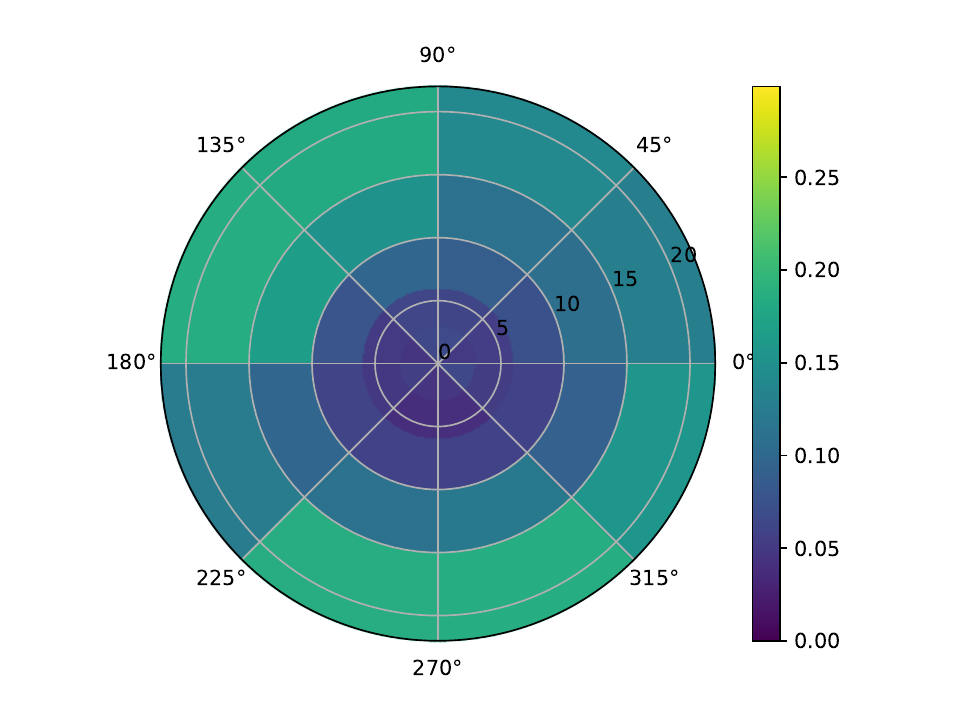}
\caption{Simulated map of limiting polarization fraction, which can be exceeded with
only $5$\% probability. The statistical correction for the effect
of noise was applied (see text).}
\label{fig:simul_pol}
\end{figure}
\begin{figure}[ht!]
\centering
\includegraphics[width=8.8cm]{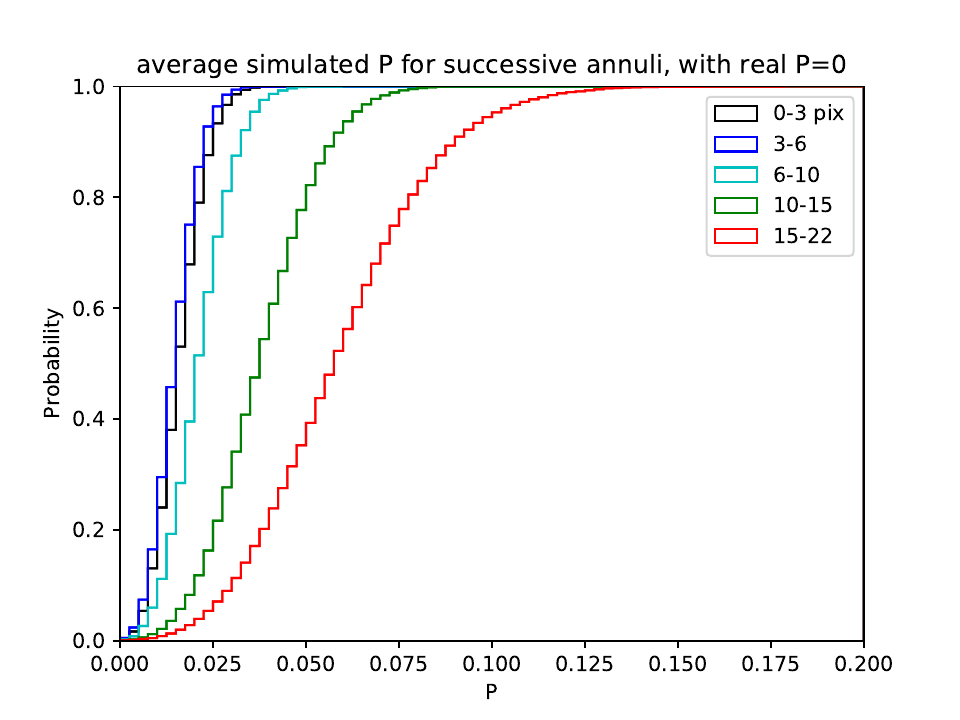}
\caption{Spurious ''observed'' polarization $P$ due to noise in successive annuli of increasing radius around
the quasar, according to simulations assuming zero real polarization. The limiting radii
(inner and outer radius) of each annulus are given in
the upper right box in pixel units; there are four pixels per arcsec, so the
outermost radius ($22$ pixels) corresponds to $5.5$\arcsec. The corresponding
cumulative probability distributions are given.}
\label{fig:simul_pol_cumul}
\end{figure}
On the eight average images (corresponding to the four position angles and the two beams, each image reaching a depth corresponding to an rms dispersion of $\sim 1.5\times 10^{-18}~\mathrm{erg}\,\mathrm{s}^{-1}\,\mathrm{cm}^{-2}\,\mathrm{arcsec}^{-2}$),
we defined a circular region of $5.5$\arcsec\ radius centered on the quasar
and divided it into five annuli and eight sectors (see Fig.~\ref{fig:P_sector}, left panel).
We summed the LAB intensity in each of the $40$ resulting areas for each of the eight images. The average polarization fraction $P$ was computed for each of the $40$ sectors;
the result is shown in Fig.~\ref{fig:P_sector}.
We used the following formulae:
\begin{equation}
P=\sqrt{\bar{Q}^2+\bar{U}^2}
\label{eq:polar_measured}
;\end{equation}
$\bar{Q}$ and $\bar{U}$ are the normalized $Q$ and $U$ Stokes parameters.
They are related to the normalized flux differences $F_\theta$ defined as
\begin{equation}
F_\theta \equiv \frac{f_\theta^{ord}-f_\theta^{ext}}{f_\theta^{ord}+f_\theta^{ext}}
\label{eq:F_formula}
\end{equation}
through the relations
\begin{eqnarray}
\bar{Q}&=&\frac{Q}{I}=\frac{F_{0.0}}{2}-\frac{F_{45.0}}{2},\\
\bar{U}&=&\frac{U}{I}=\frac{F_{22.5}}{2}-\frac{F_{67.5}}{2}~\mathrm{,}
\label{eq:QU_formulae}
\end{eqnarray}
where the subscripts refer to the HWP angles expressed in degrees.
The error on the polarization fraction obtained through propagation, under the assumption
of Gaussian shot noise, negligible readout noise, and the same background level in the
ordinary and extraordinary beams, reads
\begin{equation}
\sigma_P = \frac{1}{\sqrt{N/2}\cdot (S/N)}
\end{equation}
\citep{PR06}, where $N$ is the number of HWP angles (4 in our case), and $S/N$ is the signal-to-noise
ratio of an intensity image ($f_\theta^{ord}+f_\theta^{ext}$).
The polarization angle $\chi$ is given by the formula
\begin{equation}
\chi = \frac{1}{2}\arctan{\frac{\bar{U}}{\bar{Q}}}
,\end{equation}
and its error is given by
\begin{equation}
\sigma_\chi = \frac{1}{2\,\sqrt{N/2}\cdot P\cdot (S/N)}\equiv \frac{\sigma_P}{2\,P}
\end{equation}
\citep{PR06}.

Because of the noise and of the intrinsically positive nature of the $P$ quantity,
the values taken by the latter are strongly biased when computed in the way described
above. \citet{WK74} proposed a first-order statistical correction (their formula A3),
which in our case takes the form
\begin{equation}
P=P_m\,\sqrt{1-\left(\frac{\sigma_P}{P_m}\right)^2}
,\end{equation}
where $P$ is the corrected polarization fraction, $P_m$ is the ``measured''
polarization fraction computed through Eq.~\ref{eq:polar_measured}
and $\sigma_P$ is the rms standard deviation of $P_m$ due to background and photon
noise, computed by the standard propagation formulae. This correction is included in
the results shown in Fig.~\ref{fig:P_sector}; it is clearly not sufficient, in view
of the high $P$ value in regions with very low or negligible flux.

Averaging the polarization fraction on azimut in each annulus, we obtain the
trend shown in the left panel of Fig.~\ref{fig:Pmean}. The error bars are simply the rms standard
deviations of the eight $P$ values available in each annulus divided by $\sqrt{8}$;
they represent the error on the average polarization at given radius, under the
assumption of a normal distribution of $P$ values. Because the latter assumption
is not valid, the apparent slightly positive trend only betrays the effect of
noise, even though the latter has been partly cancelled by the correction mentioned
above. The polarization fraction being defined by construction as a positive
quantity, its frequency distribution is highly asymmetric even though the pixel
intensities follow a Gaussian distribution around the mean value on the frames.

In order to quantify the uncertainties on $P$, we produced $10\,000$ artificial
images of the LAB for each polarization angle and beam. Starting from a smoothed image
of the blob in unpolarized light (smoothed by a Gaussian PSF with $FWHM=0.75$\arcsec,
similar to the seeing of the data), we added Poisson noise corresponding to the
scatter measured on the background of the true image, and scaled it according to the
signal intensity. Zero polarization
was assumed, so that the artificial images corresponding to ordinary and extraordinary
beams, and to the various angles, differ only by the added noise (i.e., by the different
statistical realizations of the noise patterns, all realizations being drawn from
the same parent noise distribution); we also assume here that the noise depends
neither on the HWP angle nor on the beam (ordinary or extraordinary), and indeed the
measured noise is the same to an accuracy level better than 10\%. We then computed
$P$ in the same way as for the true images, including the Wardle \& Kronberg correction.
For each of the $5\times8=40$ sectors, we have determined the polarization fraction
above which only $5$\% of the random draws occur. The resulting map of this limiting
value is shown in Fig.~\ref{fig:simul_pol}; it may be considered as a map of the upper
limits to $P$ that we would obtain from ideal (i.e., free from any systematics) but
noisy data such as ours, the limit corresponding to a $95$\% probability. We note that
this map essentially reflects the surface brightness distribution of the LAB because
of the anticorrelation between the spurious signal and the signal-to-noise ratio.

For each randomized set of images, the simulated polarization fraction was averaged in
azimut in exactly the same way as for the observed data. This yields the set of five
cumulative distributions shown on Fig.~\ref{fig:simul_pol_cumul}, each distribution
corresponding to a different annulus. For example, the cumulative distribution for the
outermost annulus (red curve) shows that half of the simulated data yield a ``measured''
polarization fraction $P< 0.06$ in this annulus, while the other half yield
$0.06 < P < 0.15$, so the median $P$ value is $0.06$.
We see that both the mode
and the width of the distributions increase with radius because of increasing noise linked
with decreasing intensity. Even though the Wardle \& Kronberg correction was applied,
there is still an $\sim 5$\% probability of a
polarization fraction $P > 0.10$ in the outer annulus. This is of course a spurious, apparent
polarization due to noise alone, since zero polarization was assumed in the simulation.

The ``observed'' polarization map shown in Fig.~\ref{fig:P_sector} appears more
contrasted than the simulated one (Fig.~\ref{fig:simul_pol}) for
the $95$\% limit, which is not unexpected, but it also displays some sectors exceeding
that limit, most notably one in the outer annulus, and a few
at smaller radii.
One has to keep in mind that the simulation does not include any 
systematic error, while actually there is also some uncertainty on, for instance, the sky subtraction. Nevertheless,
the two maps are roughly and globally consistent.

In the simulation (Fig.~\ref{fig:simul_pol_cumul}), the upper limit to the azimutally
averaged $P$ (with $95$\% significance) increases from $\sim 2$\% for a projected radius
$R < 1.5$\arcsec to $\sim 10$\% at $R\simeq 4.6$\arcsec. Observations
(Fig.~\ref{fig:Pmean}) lie below or close to this limit, and noise prevents us from
determining meaningful $P$ values beyond a radius $R\sim 5\arcsec=20$~pixels. Therefore,
we conclude that our data are compatible with no polarization up to about $5$\arcsec\
or 38~kpc from the quasar.

\section{Discussion and comparison with theoretical simulations}
\begin{figure}[ht!]
\centering
\includegraphics[width=8.8cm]{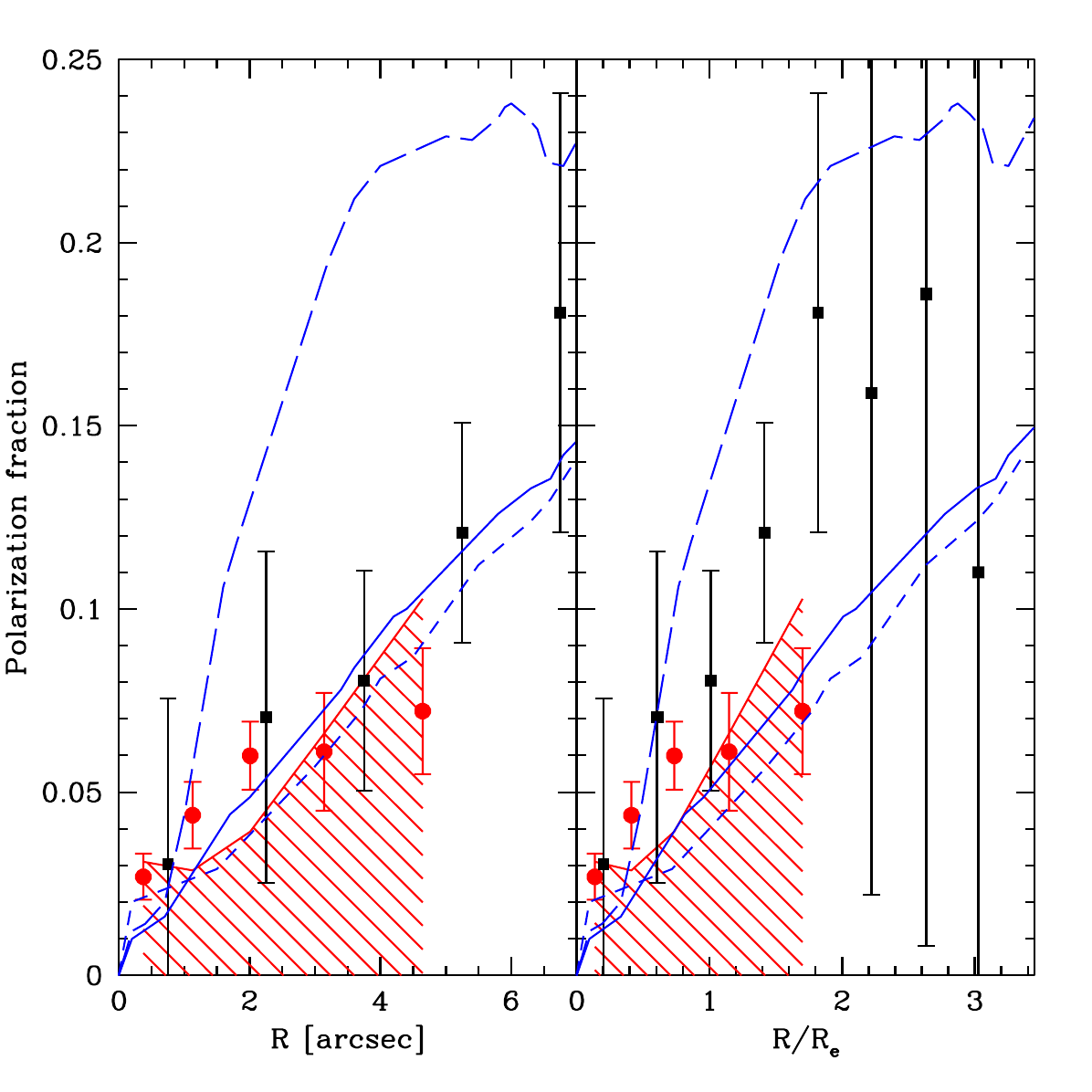}
\caption{Average polarization fraction in successive annuli as a function
of radius. {\bf Left panel:} Radius expressed in arcseconds. {\sl Red dots:} This work. The statistical correction for the effect of noise (see text) was applied.
The error bars are the rms standard deviation of the $P$ values in the  eight sectors
corresponding to each radius, divided by $\sqrt{8}$. The hatched area below the red
line represents the region in which the measured polarization fraction will
lie with a $95$\% probability if the real polarization fraction is zero. We note
that two of our measurements lie just above the red line, showing
marginally significant signal, while the others lie below the red line and are thus consistent with no polarization.
{\sl Black square dots:} Polarization fraction measured by \citet{HSS11} 
for LAB1 in the SSA22 area.
{\sl Blue lines:} Theoretical predictions by \citet{TVB16} for a synthetic LAB at
redshift $z=3$. {\sl Short-dashed line:} Median polarization profile corresponding to the
``extragalactic'' Ly$\alpha$ emission (from gas with $n_\mathrm{H}< 0.76$~cm$^{-3}$
number density).
{\sl Long-dashed line:} Same as above, but for the ``galactic'' emission
(from \ion{H}{ii} regions in the central galaxy).
{\sl Continuous line:} Same as above, but for the overall Ly$\alpha$ emission. {\bf Right panel:} Same as left panel, but with radius expressed in units or half-light radius $R_\mathrm{e}$, with $R_\mathrm{e}=2.71\,\arcsec$ for our object, $2.09\,\arcsec$ for the simulation, and $3.71\,\arcsec$ for LAB1 (see text). The three black points beyond $R/R_\mathrm{e}=2$ are omitted in the left panel because they lie beyond the limit.}
\label{fig:Pmean}
\end{figure}

Figure~\ref{fig:Pmean} also displays the polarization fraction found by \citet{HSS11}
in LAB1 (black dots). At small radii, their results are not widely different from
ours, but their object is brighter at large radii and their $P$ values are significant
there. They reach $P\simeq 0.18$ at about $6.7$\arcsec\ from the LAB center\footnote{We note that
their error bars are not defined in the same way as ours and that the red curve
relates only to our observations, not theirs.}. In our case, the flux beyond $5$\arcsec\
is too small to draw meaningful conclusions about larger radii.

The lack of significant polarization is surprising at first sight, in view of the simulations
of synthetic blobs performed, for example, by \citet{TVB16}. In their work, \citet{TVB16}
considered two distinct sources of Ly$\alpha$ photons (see Fig.~\ref{fig:Pmean} and their Fig. 5):
one related to the {\sl \emph{extragalactic}} component, made of gas with densities
$n_\mathrm{H}\leq 0.76$\,cm$^{-3}$, and one related to the {\sl \emph{galactic}} component,
in the form of \ion{H}{ii} regions lying in the central galaxy. The former component includes 
the circumgalactic gas (CGM), cold streams and other diffuse gas, while the latter
is made of galactic interstellar matter.

Based on the polarization measured by \citet{HSS11}, we would expect a polarization level similar
to that of the {\sl \emph{overall}} Ly$\alpha$ emission case (i.e., extragalactic and galactic components
combined; solid blue line in Fig.~\ref{fig:Pmean}). In such a
situation, we expect $P=10.5$\% at $R=4.6$\arcsec, while we find only $7$\%.
At this radius, the prediction coincides with our $95$\% upper limit defined by the
simulations described above; in the case of no true polarization, the measured polarization
fraction has a $95$\% probability of falling below this limit (red line in Fig.~\ref{fig:Pmean}).
\citet{TVB16} predicted $P\simeq 0.07$ at the $\sim 3.1$\arcsec radius, above our $95$\% limit;
their prediction also lies above the $95$\% limit at $2$\arcsec and  $1.1$\arcsec radii.
The only exception occurs for the innermost region, with an average radius of
$\sim 0.4$\arcsec, where we find $P\sim 0.028$, slightly above the theoretical prediction
but still below our $95$\% limit.

The short-dashed and long-dashed blue lines in Fig.~\ref{fig:Pmean} refer to the
{\sl \emph{extragalactic}} and {\sl \emph{galactic}} components of the Ly$\alpha$ emission, respectively,
that are envisaged by \citet{TVB16}.
Interestingly, the  {\sl \emph{extragalactic}} emission is extended by definition; in other words, the Ly$\alpha$ photons
are produced {\sl \emph{in situ}}, so no polarization is expected at first sight for
this component. However, the CGM density decreases outwards, so most photons are
produced near the LAB center and a significant fraction of them are scattered in the
outer regions before reaching the observer, which explains the rather high predicted polarization
fraction. Conversely, the photons corresponding to the {\sl \emph{galactic}} component are emitted
{\sl \emph{exclusively}} near the LAB center\footnote{This is because the galaxy lies near the LAB center.
If there is a merger in progress, this could change the scenario considerably, at least if the merger is in an early stage. In our object, though, the smoothness of the SB profile does not suggest any such event.}, where the galaxies lie, and many of them are
subsequently scattered toward the observer, thereby acquiring their large polarization.

The fact that the $P$ measurements fall in or near the $95$\% probability area makes any agreement with the theoretical prediction unlikely, especially as the orientation of the polarization vector
is not systematically tangential. This remains true even if we 
consider only the extragalactic component (short-dashed curve in 
Fig.~\ref{fig:Pmean}). This is even more striking if we consider the 
galactic component alone (long-dashed curve), that is, assuming no {\sl \emph{in 
situ}} extended emission. The observed extended emission can clearly not 
be due to the scattering of  Ly$\alpha$ photons emitted by the quasar, 
since the polarization would then exceed $20$\% in the outer regions.

In this comparison with the simulated LAB, one has to keep in mind that 
the latter differs in some important respects from the real observed 
LAB, even though the redshifts are
roughly the same. This holds not only for our object, but also for LAB1 
observed by \citet{HSS11}. The simulated LAB is much less luminous, 
with an SB about five times lower than our observed one out to radii of 
about $5-6$\arcsec. Thus, if the polarization properties are tied to 
the gas distribution, comparing them at the same physical radius for 
widely different LABs may not be optimal. A rather natural way to 
compare different LABs is to plot $P$ as a function of a radius 
normalized to the effective radii $R_\mathrm{e}$ of the respective LABs 
(where $R_\mathrm{e}$ is the radius encompassing half of the LAB total 
luminosity). We estimate $R_\mathrm{e}=2.79\,\arcsec$ for 
SDSS~J1240+1455, $3.71\,\arcsec$ for LAB1, and $2.09\,\arcsec$ for the 
simulated LAB; the polarization fraction is plotted as a function of 
the ratio $R/R_\mathrm{e}$ in the right panel of Fig.~\ref{fig:Pmean}. 
Because the effective radius of our LAB is not very far from that 
of the simulated one, little has changed compared to the left panel of 
Fig.~\ref{fig:Pmean}; the simulated overall Ly$\alpha$ emission is
now slightly below the observed $P$ upper limit (red line).
Conversely, all LAB1 data points \citep{HSS11} now lie above the curve 
of the simulated LAB (continuous blue line), while they where fully 
compatible with it in the left panel of Fig.~\ref{fig:Pmean}. They lie 
roughly halfway between the so-called galactic and overall emission 
curves, suggesting a larger contribution, in LAB1, of the galactic 
Ly$\alpha$ emission than of the extragalactic emission to the 
polarization.

One explanation for the difference between LAB1 and the simulation is 
the steeper SB profile (or smaller $R_\mathrm{e}$) of the latter. This 
steeper profile may be linked to the fact that the underlying 
cosmological simulation lacks feedback and metals, leading to a dearth 
of gas (including cold gas) in the CGM and eventually to an 
underestimated $R_\mathrm{e}$ relative to a more realistic model. This 
would probably affect the polarization profile too.

Our object differs greatly from the simulation by the presence of a 
quasar; there is no AGN activity whatsoever in the simulation, and the 
ionizing radiation only comes from the UV background. This would also 
partly affect the radial \ion{H}{i} profile, and consequently the 
Ly$\alpha$ and polarization profiles.
Thus, we can interpret the observed polarization (or lack of it) in several ways. The gas
may be largely photoionized by the quasar, so the Ly$\alpha$ emission may be essentially due to spontaneous emission following recombination, rather than
to scattering of Ly$\alpha$ photons. Indeed, photoionization reduces the \ion{H}{i} number density, hence the number of potential scatterers; thus, two mechanisms concur here to reduce polarization, compared to the simulated LAB that hosts no quasar. Another possibility is that, even if scattering
does occur, the scattering angle never takes the $\pi/2$ value because the gas
lies in front of, or behind the quasar instead of lying in the same plane perpendicular
to the line of sight.

\section{Conclusion}
We present a detailed analysis of the LAB surrounding the radio-quiet quasar
SDSS~J1240+1455 using narrow-band imaging polarimetry. Taking advantage of the PDLA
acting as a natural coronagraph, we were able to describe the morphology of the LAB and
put an upper limit on its polarization fraction. This is the first polarization study
of a LAB obviously hosting a radio-quiet quasar.

The main results can be summarized as follows.
\begin{itemize}
\item The LAB shows a complex filamentary structure, which is typical of other LABs
and that would be worth exploring further with an IFU instrument such as MUSE at VLT.
\item The LAB structure could be characterized down to its center, taking advantage
of the coronagraphic effect provided by the PDLA. However, the filter was still too
wide to completely exclude any light from the quasar, and subtracting this light
was necessary, introducing some uncertainty. A bimodal structure
was found in the LAB center that may be due either to a foreground 
absorption possibly linked to the PDLA or to some density fluctuation 
in the LAB itself. Here too, MUSE observations may shed light on the 
nature of this feature.
\item The azimuthally averaged surface brightness of the LAB decreases 
exponentially from its center, but with a steeper slope in the inner 
region than in the outer one. 
\item No significant linear polarization could be seen in the LAB 
within $5.5$\arcsec\ of the quasar, in any of the $40$ sectors 
explored. Likewise, averaging the polarization fraction in concentric 
annuli provided only upper limits to it. The polarization angles 
show no significant trend toward tangential orientation.
Our data are compatible with zero polarization.
\item The lack of polarization we find in this LAB is not compatible with the
properties of the synthetic LAB modeled by \citet{TVB16}, suggesting that the main
Ly$\alpha$ emission mechanism is {\sl \emph{in situ}} recombination in gas photoionized
by the central quasar.
\end{itemize}
More observation with, for instance, the MUSE integral field unit spectrograph would be most
interesting to study the morphology of the inner part of the LAB, as well as the
excitation and ionization properties of the gas and its kinematics. In particular,
the \ion{C}{iv}~$\lambda 1548$ and \ion{He}{ii}~$\lambda 1640$ lines would bring
interesting information in this regard. The extension
of the LAB may also prove wider than that found here.
This study also calls for improving simulations of the extended (polarized) Lyman-alpha emission with more detailed modeling, including, in particular, the effect of photoionization by an AGN.

\begin{acknowledgements}
A.V. acknowledges support from the SNF Professorship PP00P2\_176808 and the ERC starting grant ERC-757258-TRIPLE.
MT acknowledges support from the NWO grant 016.VIDI.189.162 (“ODIN”).
D.C.'s work has been partially supported by grants from the German Research Foundation (HA3555-14/1, CH71-34-3) and the Israeli Science Foundation (2398/19).
P.N. thanks Dr. Henri Boffin and the ESO staff at Paranal for the 
transmission curve of the OIII+50 interference filter in the SR mode, Dr. Sabine Moehler (ESO) for clarifying a technical point, and Dr. Thibault North
for his help with the Python language.
We also thank Dr. Dominique Sluse for constructive remarks on early
versions of the draft, and the anonymous referee for very useful
remarks.
\end{acknowledgements}

\bibliographystyle{aa}
\bibliography{agn}

\begin{thebibliography}{97}
\expandafter\ifx\csname natexlab\endcsname\relax\def\natexlab#1{#1}\fi

\bibitem[{{Appenzeller} {et~al.}(1998){Appenzeller}, {Fricke}, {F{\"u}rtig},
  {G{\"a}ssler}, {H{\"a}fner}, {Harke}, {Hess}, {Hummel}, {J{\"u}rgens},
  {Kudritzki}, {Mantel}, {Meisl}, {Muschielok}, {Nicklas}, {Rupprecht},
  {Seifert}, {Stahl}, {Szeifert}, \& {Tarantik}}]{AFF98}
{Appenzeller}, I., {Fricke}, K., {F{\"u}rtig}, W., {et~al.} 1998, The
  Messenger, 94, 1

\bibitem[{{Beck} {et~al.}(2016){Beck}, {Scarlata}, {Hayes}, {Dijkstra}, \&
  {Jones}}]{BSH16}
{Beck}, M., {Scarlata}, C., {Hayes}, M., {Dijkstra}, M., \& {Jones}, T.~J.
  2016, \apj, 818, 138

\bibitem[{{Binney} \& {Tremaine}(2008)}]{BT08}
{Binney}, J. \& {Tremaine}, S. 2008, {Galactic Dynamics: Second Edition}
  (Princeton University Press)

\bibitem[{{Boffin} {et~al.}(2013){Boffin}, {Dumas}, \& {Kaufer}}]{BDK13}
{Boffin}, H., {Dumas}, C., \& {Kaufer}, A. 2013, VLT-MAN-ESO-13100-1543, 92.0,
  535

\bibitem[{{Borisova} {et~al.}(2016){Borisova}, {Cantalupo}, {Lilly}, {Marino},
  {Gallego}, {Bacon}, {Blaizot}, {Bouch{\'e}}, {Brinchmann}, {Carollo},
  {Caruana}, {Finley}, {Herenz}, {Richard}, {Schaye}, {Straka}, {Turner},
  {Urrutia}, {Verhamme}, \& {Wisotzki}}]{BCL16}
{Borisova}, E., {Cantalupo}, S., {Lilly}, S.~J., {et~al.} 2016, \apj, 831, 39

\bibitem[{{Bremer} {et~al.}(1992){Bremer}, {Fabian}, {Sargent}, {Steidel},
  {Boksenberg}, \& {Johnstone}}]{bremer92}
{Bremer}, M.~N., {Fabian}, A.~C., {Sargent}, W.~L.~W., {et~al.} 1992, \mnras,
  258, 23

\bibitem[{{Bunker} {et~al.}(2003){Bunker}, {Smith}, {Spinrad}, {Stern}, \&
  {Warren}}]{BSS03}
{Bunker}, A., {Smith}, J., {Spinrad}, H., {Stern}, D., \& {Warren}, S. 2003,
  \apss, 284, 357

\bibitem[{{Cabot} {et~al.}(2016){Cabot}, {Cen}, \& {Zheng}}]{CCZ16}
{Cabot}, S.~H.~C., {Cen}, R., \& {Zheng}, Z. 2016, \mnras, 462, 1076

\bibitem[{{Caminha} {et~al.}(2016){Caminha}, {Karman}, {Rosati}, {Caputi},
  {Arrigoni Battaia}, {Balestra}, {Grillo}, {Mercurio}, {Nonino}, \&
  {Vanzella}}]{CKR16}
{Caminha}, G.~B., {Karman}, W., {Rosati}, P., {et~al.} 2016, \aap, 595, A100

\bibitem[{{Cantalupo} {et~al.}(2014){Cantalupo}, {Arrigoni-Battaia},
  {Prochaska}, {Hennawi}, \& {Madau}}]{CAP14}
{Cantalupo}, S., {Arrigoni-Battaia}, F., {Prochaska}, J.~X., {Hennawi}, J.~F.,
  \& {Madau}, P. 2014, \nat, 506, 63

\bibitem[{{Cantalupo} {et~al.}(2012){Cantalupo}, {Lilly}, \&
  {Haehnelt}}]{CLH12}
{Cantalupo}, S., {Lilly}, S.~J., \& {Haehnelt}, M.~G. 2012, \mnras, 425, 1992

\bibitem[{{Cantalupo} {et~al.}(2007){Cantalupo}, {Lilly}, \&
  {Porciani}}]{CLP07}
{Cantalupo}, S., {Lilly}, S.~J., \& {Porciani}, C. 2007, \apj, 657, 135

\bibitem[{{Chapman} {et~al.}(2005){Chapman}, {Blain}, {Smail}, \&
  {Ivison}}]{CBS05}
{Chapman}, S.~C., {Blain}, A.~W., {Smail}, I., \& {Ivison}, R.~J. 2005, \apj,
  622, 772

\bibitem[{{Chelouche} {et~al.}(2008){Chelouche}, {M{\'e}nard}, {Bowen}, \&
  {Gnat}}]{CMBG08}
{Chelouche}, D., {M{\'e}nard}, B., {Bowen}, D.~V., \& {Gnat}, O. 2008, \apj,
  683, 55

\bibitem[{{Christensen} {et~al.}(2006){Christensen}, {Jahnke}, {Wisotzki}, \&
  {S\'anchez}}]{CJW06}
{Christensen}, L., {Jahnke}, K., {Wisotzki}, L., \& {S\'anchez}, S.~F. 2006,
  \aap, 459, 717

\bibitem[{{Cicone} {et~al.}(2015){Cicone}, {Maiolino}, {Gallerani}, {Neri},
  {Ferrara}, {Sturm}, {Fiore}, {Piconcelli}, \& {Feruglio}}]{CMG15}
{Cicone}, C., {Maiolino}, R., {Gallerani}, S., {et~al.} 2015, \aap, 574, A14

\bibitem[{{Cikota} {et~al.}(2016){Cikota}, {Patat}, {Cikota}, \&
  {Faran}}]{CPC16}
{Cikota}, A., {Patat}, F., {Cikota}, S., \& {Faran}, T. 2016, \mnras
  [\eprint[arXiv]{1610.00722}]

\bibitem[{{Cimatti} {et~al.}(1998){Cimatti}, {di Serego Alighieri}, {Vernet},
  {Cohen}, \& {Fosbury}}]{CdV98}
{Cimatti}, A., {di Serego Alighieri}, S., {Vernet}, J., {Cohen}, M.~H., \&
  {Fosbury}, R.~A.~E. 1998, \apjl, 499, L21

\bibitem[{{Ciotti} \& {Bertin}(1999)}]{CB99}
{Ciotti}, L. \& {Bertin}, G. 1999, \aap, 352, 447

\bibitem[{{Daddi} {et~al.}(2022){Daddi}, {Rich}, {Valentino}, {Jin},
  {Delvecchio}, {Liu}, {Strazzullo}, {Neill}, {Gobat}, {Finoguenov},
  {Bournaud}, {Elbaz}, {Kalita}, {O'Sullivan}, \& {Wang}}]{DRV22}
{Daddi}, E., {Rich}, R.~M., {Valentino}, F., {et~al.} 2022, \apjl, 926, L21

\bibitem[{{Dey} {et~al.}(2005){Dey}, {Bian}, {Soifer}, {Brand}, {Brown},
  {Chaffee}, {Le Floc'h}, {Hill}, {Houck}, {Jannuzi}, {Rieke}, {Weedman},
  {Brodwin}, \& {Eisenhardt}}]{DBS05}
{Dey}, A., {Bian}, C., {Soifer}, B.~T., {et~al.} 2005, \apj, 629, 654

\bibitem[{{Dijkstra} \& {Loeb}(2008)}]{DL08}
{Dijkstra}, M. \& {Loeb}, A. 2008, \mnras, 386, 492

\bibitem[{{Dijkstra} \& {Loeb}(2009)}]{DL09}
{Dijkstra}, M. \& {Loeb}, A. 2009, \mnras, 400, 1109

\bibitem[{{Falomo} {et~al.}(2014){Falomo}, {Bettoni}, {Karhunen}, {Kotilainen},
  \& {Uslenghi}}]{FBK14}
{Falomo}, R., {Bettoni}, D., {Karhunen}, K., {Kotilainen}, J.~K., \&
  {Uslenghi}, M. 2014, \mnras, 440, 476

\bibitem[{{Fardal} {et~al.}(2001){Fardal}, {Katz}, {Gardner}, {Hernquist},
  {Weinberg}, \& {Dav{\'e}}}]{FKG01}
{Fardal}, M.~A., {Katz}, N., {Gardner}, J.~P., {et~al.} 2001, \apj, 562, 605

\bibitem[{{Fathivavsari} {et~al.}(2016){Fathivavsari}, {Petitjean},
  {Noterdaeme}, {P{\^a}ris}, {Finley}, {L{\'o}pez}, \& {Srianand}}]{FPN16}
{Fathivavsari}, H., {Petitjean}, P., {Noterdaeme}, P., {et~al.} 2016, \mnras,
  461, 1816

\bibitem[{{Fossati} {et~al.}(2007){Fossati}, {Bagnulo}, {Mason}, \& {Landi
  Degl'Innocenti}}]{FBM07}
{Fossati}, L., {Bagnulo}, S., {Mason}, E., \& {Landi Degl'Innocenti}, E. 2007,
  in Astronomical Society of the Pacific Conference Series, Vol. 364, The
  Future of Photometric, Spectrophotometric and Polarimetric Standardization,
  ed. C.~{Sterken}, 503

\bibitem[{{Francis} {et~al.}(2013){Francis}, {Dopita}, {Colbert}, {Palunas},
  {Scarlata}, {Teplitz}, {Williger}, \& {Woodgate}}]{FDC13}
{Francis}, P.~J., {Dopita}, M.~A., {Colbert}, J.~W., {et~al.} 2013, \mnras,
  428, 28

\bibitem[{{Francis} \& {McDonnell}(2006)}]{FMD06}
{Francis}, P.~J. \& {McDonnell}, S. 2006, \mnras, 370, 1372

\bibitem[{{Fynbo} {et~al.}(1999){Fynbo}, {M{\o}ller}, \& {Warren}}]{FMW99}
{Fynbo}, J.~U., {M{\o}ller}, P., \& {Warren}, S.~J. 1999, \mnras, 305, 849

\bibitem[{{Geach} {et~al.}(2009){Geach}, {Alexander}, {Lehmer}, {Smail},
  {Matsuda}, {Chapman}, {Scharf}, {Ivison}, {Volonteri}, {Yamada}, {Blain},
  {Bower}, {Bauer}, \& {Basu-Zych}}]{GALS09}
{Geach}, J.~E., {Alexander}, D.~M., {Lehmer}, B.~D., {et~al.} 2009, \apj, 700,
  1

\bibitem[{{Geach} {et~al.}(2014){Geach}, {Bower}, {Alexander}, {Blain},
  {Bremer}, {Chapin}, {Chapman}, {Clements}, {Coppin}, {Dunlop}, {Farrah},
  {Jenness}, {Koprowski}, {Micha{\l}owski}, {Robson}, {Scott}, {Smith},
  {Spaans}, {Swinbank}, \& {van der Werf}}]{GBA14}
{Geach}, J.~E., {Bower}, R.~G., {Alexander}, D.~M., {et~al.} 2014, \apj, 793,
  22

\bibitem[{{Geach} {et~al.}(2016){Geach}, {Narayanan}, {Matsuda}, {Hayes},
  {Mas-Ribas}, {Dijkstra}, {Steidel}, {Chapman}, {Feldmann}, {Avison},
  {Agertz}, {Ao}, {Birkinshaw}, {Bremer}, {Clements}, {Dannerbauer}, {Farrah},
  {Harrison}, {Kubo}, {Micha{\l}owski}, {Scott}, {Smith}, {Spaans}, {Simpson},
  {Swinbank}, {Taniguchi}, {van der Werf}, {Verma}, \& {Yamada}}]{GNM16}
{Geach}, J.~E., {Narayanan}, D., {Matsuda}, Y., {et~al.} 2016, \apj, 832, 37

\bibitem[{{Goto} {et~al.}(2012){Goto}, {Utsumi}, {Walsh}, {Hattori},
  {Miyazaki}, \& {Yamauchi}}]{GUW12}
{Goto}, T., {Utsumi}, Y., {Walsh}, J.~R., {et~al.} 2012, \mnras, 421, L77

\bibitem[{{Haiman} \& {Rees}(2001)}]{HR01}
{Haiman}, Z. \& {Rees}, M.~J. 2001, \apj, 556, 87

\bibitem[{{Haiman} {et~al.}(2000){Haiman}, {Spaans}, \& {Quataert}}]{HSQ00}
{Haiman}, Z., {Spaans}, M., \& {Quataert}, E. 2000, \apjl, 537, L5

\bibitem[{{Hamilton} {et~al.}(2002){Hamilton}, {Casertano}, \&
  {Turnshek}}]{HCT02}
{Hamilton}, T.~S., {Casertano}, S., \& {Turnshek}, D.~A. 2002, \apj, 576, 61

\bibitem[{{Hamuy} {et~al.}(1992){Hamuy}, {Walker}, {Suntzeff}, {Gigoux},
  {Heathcote}, \& {Phillips}}]{HWS92}
{Hamuy}, M., {Walker}, A.~R., {Suntzeff}, N.~B., {et~al.} 1992, \pasp, 104, 533

\bibitem[{{Hayes} {et~al.}(2011){Hayes}, {Scarlata}, \& {Siana}}]{HSS11}
{Hayes}, M., {Scarlata}, C., \& {Siana}, B. 2011, \nat, 476, 304

\bibitem[{{Heckman} {et~al.}(1991){Heckman}, {Lehnert}, {Miley}, \& {van
  Breugel}}]{H91a}
{Heckman}, T.~M., {Lehnert}, M.~D., {Miley}, G.~K., \& {van Breugel}, W. 1991,
  \apj, 381, 373

\bibitem[{{Hennawi} {et~al.}(2009){Hennawi}, {Prochaska}, {Kollmeier}, \&
  {Zheng}}]{HPKZ09}
{Hennawi}, J.~F., {Prochaska}, J.~X., {Kollmeier}, J., \& {Zheng}, Z. 2009,
  \apjl, 693, L49

\bibitem[{{Hu} {et~al.}(1996){Hu}, {McMahon}, \& {Egami}}]{HME96}
{Hu}, E.~M., {McMahon}, R.~G., \& {Egami}, E. 1996, \apjl, 459, L53

\bibitem[{{Humphrey} {et~al.}(2013){Humphrey}, {Vernet},
  {Villar-Mart{\'{\i}}n}, {di Serego Alighieri}, {Fosbury}, \&
  {Cimatti}}]{HVV13}
{Humphrey}, A., {Vernet}, J., {Villar-Mart{\'{\i}}n}, M., {et~al.} 2013, \apjl,
  768, L3

\bibitem[{{Hutchings} {et~al.}(1984){Hutchings}, {Crampton}, \&
  {Campbell}}]{HCC84}
{Hutchings}, J.~B., {Crampton}, D., \& {Campbell}, B. 1984, \apj, 280, 41

\bibitem[{{Hutsem{\'e}kers} {et~al.}(2014){Hutsem{\'e}kers}, {Braibant},
  {Pelgrims}, \& {Sluse}}]{HBP14}
{Hutsem{\'e}kers}, D., {Braibant}, L., {Pelgrims}, V., \& {Sluse}, D. 2014,
  \aap, 572, A18

\bibitem[{{Hyv{\"o}nen} {et~al.}(2007){Hyv{\"o}nen}, {Kotilainen},
  {{\"O}rndahl}, {Falomo}, \& {Uslenghi}}]{HKO07}
{Hyv{\"o}nen}, T., {Kotilainen}, J.~K., {{\"O}rndahl}, E., {Falomo}, R., \&
  {Uslenghi}, M. 2007, \aap, 462, 525

\bibitem[{{Jiang} {et~al.}(2016){Jiang}, {Zhou}, {Pan}, {Jiang}, {Shu}, {Wang},
  {Gu}, {Li}, {Wu}, {Shi}, {Ji}, {Tian}, \& {Zhang}}]{JZP16}
{Jiang}, P., {Zhou}, H., {Pan}, X., {et~al.} 2016, \apj, 821, 1

\bibitem[{{Kashikawa} {et~al.}(2014){Kashikawa}, {Misawa}, {Minowa}, {Okoshi},
  {Hattori}, {Toshikawa}, {Ishikawa}, \& {Onoue}}]{KMM14}
{Kashikawa}, N., {Misawa}, T., {Minowa}, Y., {et~al.} 2014, \apj, 780, 116

\bibitem[{{Kawamuro} {et~al.}(2017){Kawamuro}, {Schirmer}, {Turner}, {Davies},
  \& {Ichikawa}}]{KST17}
{Kawamuro}, T., {Schirmer}, M., {Turner}, J.~E.~H., {Davies}, R.~L., \&
  {Ichikawa}, K. 2017, \apj, 848, 42

\bibitem[{{Kim} {et~al.}(2020){Kim}, {Yang}, {Zabludoff}, {Smith}, {Jannuzi},
  {Lee}, {Hwang}, \& {Park}}]{KYZ20}
{Kim}, E., {Yang}, Y., {Zabludoff}, A., {et~al.} 2020, \apj, 894, 33

\bibitem[{{Kim} {et~al.}(2023){Kim}, {Yang}, {Zabludoff}, {Smith}, {Jannuzi},
  {Lee}, {Hwang}, \& {Park}}]{KYZ23}
{Kim}, E., {Yang}, Y., {Zabludoff}, A., {et~al.} 2023, in American Astronomical
  Society Meeting Abstracts, Vol.~55, American Astronomical Society Meeting
  Abstracts, 460.15

\bibitem[{{Leclercq} {et~al.}(2017){Leclercq}, {Bacon}, {Wisotzki}, {Mitchell},
  {Garel}, {Verhamme}, {Blaizot}, {Hashimoto}, {Herenz}, {Conseil},
  {Cantalupo}, {Inami}, {Contini}, {Richard}, {Maseda}, {Schaye}, {Marino},
  {Akhlaghi}, {Brinchmann}, \& {Carollo}}]{LBW17}
{Leclercq}, F., {Bacon}, R., {Wisotzki}, L., {et~al.} 2017, ArXiv e-prints
  [\eprint[arXiv]{1710.10271}]

\bibitem[{{Lehnert} \& {Becker}(1998)}]{LB98}
{Lehnert}, M.~D. \& {Becker}, R.~H. 1998, \aap, 332, 514

\bibitem[{{Luo} {et~al.}(2021){Luo}, {Heckman}, {Hwang}, {Rowlands},
  {S{\'a}nchez-Menguiano}, {Riffel}, {Bizyaev}, {Andrews},
  {Fern{\'a}ndez-Trincado}, {Drory}, {S{\'a}nchez Almeida}, {Maiolino}, {Lane},
  \& {Argudo-Fern{\'a}ndez}}]{LHH21}
{Luo}, Y., {Heckman}, T., {Hwang}, H.-C., {et~al.} 2021, \apj, 908, 183

\bibitem[{{Martin} {et~al.}(2014{\natexlab{a}}){Martin}, {Chang},
  {Matuszewski}, {Morrissey}, {Rahman}, {Moore}, \& {Steidel}}]{MCM14a}
{Martin}, D.~C., {Chang}, D., {Matuszewski}, M., {et~al.} 2014{\natexlab{a}},
  \apj, 786, 106

\bibitem[{{Martin} {et~al.}(2014{\natexlab{b}}){Martin}, {Chang},
  {Matuszewski}, {Morrissey}, {Rahman}, {Moore}, {Steidel}, \&
  {Matsuda}}]{MCM14b}
{Martin}, D.~C., {Chang}, D., {Matuszewski}, M., {et~al.} 2014{\natexlab{b}},
  \apj, 786, 107

\bibitem[{{Martin} {et~al.}(2015){Martin}, {Matuszewski}, {Morrissey}, {Neill},
  {Moore}, {Cantalupo}, {Prochaska}, \& {Chang}}]{MMM15}
{Martin}, D.~C., {Matuszewski}, M., {Morrissey}, P., {et~al.} 2015, \nat, 524,
  192

\bibitem[{{Matsuda} {et~al.}(2009){Matsuda}, {Nakamura}, {Morimoto}, {Smail},
  {De Breuck}, {Ohta}, {Kodama}, {Inoue}, {Hayashino}, {Kousai}, {Nakamura},
  {Horie}, {Yamada}, {Kitamura}, {Saito}, {Taniguchi}, {Tanaka}, \&
  {Hibon}}]{MNM09}
{Matsuda}, Y., {Nakamura}, Y., {Morimoto}, N., {et~al.} 2009, \mnras, 400, L66

\bibitem[{{Matsuda} {et~al.}(2004){Matsuda}, {Yamada}, {Hayashino}, {Tamura},
  {Yamauchi}, {Ajiki}, {Fujita}, {Murayama}, {Nagao}, {Ohta}, {Okamura},
  {Ouchi}, {Shimasaku}, {Shioya}, \& {Taniguchi}}]{MYHT04}
{Matsuda}, Y., {Yamada}, T., {Hayashino}, T., {et~al.} 2004, \aj, 128, 569

\bibitem[{{Matsuda} {et~al.}(2011){Matsuda}, {Yamada}, {Hayashino}, {Yamauchi},
  {Nakamura}, {Morimoto}, {Ouchi}, {Ono}, {Kousai}, {Nakamura}, {Horie},
  {Fujii}, {Umemura}, \& {Mori}}]{MYH11}
{Matsuda}, Y., {Yamada}, T., {Hayashino}, T., {et~al.} 2011, \mnras, 410, L13

\bibitem[{{Michalewicz} {et~al.}(2023){Michalewicz}, {Millon}, {Dux}, \&
  {Courbin}}]{Michalewicz2023}
{Michalewicz}, K., {Millon}, M., {Dux}, F., \& {Courbin}, F. 2023, The Journal
  of Open Source Software, 8, 5340

\bibitem[{{M{\o}ller} {et~al.}(1998){M{\o}ller}, {Warren}, \& {Fynbo}}]{MWF98}
{M{\o}ller}, P., {Warren}, S.~J., \& {Fynbo}, J.~U. 1998, \aap, 330, 19

\bibitem[{{Nilsson} {et~al.}(2006){Nilsson}, {Fynbo}, {M{\o}ller},
  {Sommer-Larsen}, \& {Ledoux}}]{NFM06}
{Nilsson}, K.~K., {Fynbo}, J.~P.~U., {M{\o}ller}, P., {Sommer-Larsen}, J., \&
  {Ledoux}, C. 2006, \aap, 452, L23

\bibitem[{{North} {et~al.}(2012){North}, {Courbin}, {Eigenbrod}, \&
  {Chelouche}}]{NCE12}
{North}, P.~L., {Courbin}, F., {Eigenbrod}, A., \& {Chelouche}, D. 2012, \aap,
  542, A91

\bibitem[{{Nyland} {et~al.}(2020){Nyland}, {Dong}, {Patil}, {Lacy}, {van
  Velzen}, {Kimball}, {Sarbadhicary}, {Hallinan}, {Baldassare}, {Clarke},
  {Goulding}, {Greene}, {Hughes}, {Kassim}, {Kunert-Bajraszewska}, {Maccarone},
  {Mooley}, {Mukherjee}, {Peters}, {Petrov}, {Polisensky}, {Rujopakarn},
  {Whittle}, \& {Vaccari}}]{NDP20}
{Nyland}, K., {Dong}, D.~Z., {Patil}, P., {et~al.} 2020, \apj, 905, 74

\bibitem[{{Overzier} {et~al.}(2013){Overzier}, {Nesvadba}, {Dijkstra}, {Hatch},
  {Lehnert}, {Villar-Mart{\'\i}n}, {Wilman}, \& {Zirm}}]{OND13}
{Overzier}, R.~A., {Nesvadba}, N.~P.~H., {Dijkstra}, M., {et~al.} 2013, \apj,
  771, 89

\bibitem[{{Patat} {et~al.}(2011){Patat}, {Moehler}, {O'Brien}, {Pompei},
  {Bensby}, {Carraro}, {de Ugarte Postigo}, {Fox}, {Gavignaud}, {James},
  {Korhonen}, {Ledoux}, {Randall}, {Sana}, {Smoker}, {Stefl}, \&
  {Szeifert}}]{PMO11}
{Patat}, F., {Moehler}, S., {O'Brien}, K., {et~al.} 2011, \aap, 527, A91

\bibitem[{{Patat} \& {Romaniello}(2006)}]{PR06}
{Patat}, F. \& {Romaniello}, M. 2006, \pasp, 118, 146

\bibitem[{{Percival} {et~al.}(2001){Percival}, {Miller}, {McLure}, \&
  {Dunlop}}]{PMM01}
{Percival}, W.~J., {Miller}, L., {McLure}, R.~J., \& {Dunlop}, J.~S. 2001,
  \mnras, 322, 843

\bibitem[{{Prescott} {et~al.}(2015){Prescott}, {Momcheva}, {Brammer}, {Fynbo},
  \& {M{\o}ller}}]{PMB15}
{Prescott}, M.~K.~M., {Momcheva}, I., {Brammer}, G.~B., {Fynbo}, J.~P.~U., \&
  {M{\o}ller}, P. 2015, \apj, 802, 32

\bibitem[{{Prescott} {et~al.}(2011){Prescott}, {Smith}, {Schmidt}, \&
  {Dey}}]{PSS11}
{Prescott}, M.~K.~M., {Smith}, P.~S., {Schmidt}, G.~D., \& {Dey}, A. 2011,
  \apjl, 730, L25

\bibitem[{{Rauch} {et~al.}(2013){Rauch}, {Becker}, {Haehnelt}, {Carswell}, \&
  {Gauthier}}]{RBH13}
{Rauch}, M., {Becker}, G.~D., {Haehnelt}, M.~G., {Carswell}, R.~F., \&
  {Gauthier}, J.-R. 2013, \mnras, 431, L68

\bibitem[{{Rauch} {et~al.}(2011){Rauch}, {Becker}, {Haehnelt}, {Gauthier},
  {Ravindranath}, \& {Sargent}}]{RBH11}
{Rauch}, M., {Becker}, G.~D., {Haehnelt}, M.~G., {et~al.} 2011, \mnras, 418,
  1115

\bibitem[{{Rosdahl} \& {Blaizot}(2012)}]{RB12}
{Rosdahl}, J. \& {Blaizot}, J. 2012, \mnras, 423, 344

\bibitem[{{Sanderson} {et~al.}(2021){Sanderson}, {Prescott}, {Christensen},
  {Fynbo}, \& {M{\o}ller}}]{SPC21}
{Sanderson}, K.~N., {Prescott}, M. K.~M., {Christensen}, L., {Fynbo}, J., \&
  {M{\o}ller}, P. 2021, \apj, 923, 252

\bibitem[{{Scarlata} {et~al.}(2009){Scarlata}, {Colbert}, {Teplitz}, {Bridge},
  {Francis}, {Palunas}, {Siana}, {Williger}, \& {Woodgate}}]{SCT09}
{Scarlata}, C., {Colbert}, J., {Teplitz}, H.~I., {et~al.} 2009, \apj, 706, 1241

\bibitem[{{Schirmer} {et~al.}(2016){Schirmer}, {Malhotra}, {Levenson}, {Fu},
  {Davies}, {Keel}, {Torrey}, {Bennert}, {Pancoast}, \& {Turner}}]{SML16}
{Schirmer}, M., {Malhotra}, S., {Levenson}, N.~A., {et~al.} 2016, \mnras, 463,
  1554

\bibitem[{{Smailagic} {et~al.}(2017){Smailagic}, {Micic}, {Bogosavljevic}, \&
  {Martinovic}}]{SMB17}
{Smailagic}, M., {Micic}, M., {Bogosavljevic}, M., \& {Martinovic}, N. 2017,
  Publications de l'Observatoire Astronomique de Beograd, 96, 287

\bibitem[{{Smith} \& {Jarvis}(2007)}]{SJ07}
{Smith}, D.~J.~B. \& {Jarvis}, M.~J. 2007, \mnras, 378, L49

\bibitem[{{Smith} {et~al.}(2009){Smith}, {Jarvis}, {Simpson}, \&
  {Mart{\'{\i}}nez-Sansigre}}]{SJS09}
{Smith}, D.~J.~B., {Jarvis}, M.~J., {Simpson}, C., \&
  {Mart{\'{\i}}nez-Sansigre}, A. 2009, \mnras, 393, 309

\bibitem[{{Smith} {et~al.}(1986){Smith}, {Heckman}, {Bothun}, {Romanishin}, \&
  {Balick}}]{SHB86}
{Smith}, E.~P., {Heckman}, T.~M., {Bothun}, G.~D., {Romanishin}, W., \&
  {Balick}, B. 1986, \apj, 306, 64

\bibitem[{{Steidel} {et~al.}(2000){Steidel}, {Adelberger}, {Shapley},
  {Pettini}, {Dickinson}, \& {Giavalisco}}]{SAS00}
{Steidel}, C.~C., {Adelberger}, K.~L., {Shapley}, A.~E., {et~al.} 2000, \apj,
  532, 170

\bibitem[{{Steidel} {et~al.}(2011){Steidel}, {Bogosavljevi{\'c}}, {Shapley},
  {Kollmeier}, {Reddy}, {Erb}, \& {Pettini}}]{SBS11}
{Steidel}, C.~C., {Bogosavljevi{\'c}}, M., {Shapley}, A.~E., {et~al.} 2011,
  \apj, 736, 160

\bibitem[{{Trebitsch} {et~al.}(2016){Trebitsch}, {Verhamme}, {Blaizot}, \&
  {Rosdahl}}]{TVB16}
{Trebitsch}, M., {Verhamme}, A., {Blaizot}, J., \& {Rosdahl}, J. 2016, \aap,
  593, A122

\bibitem[{{Vernet} {et~al.}(1999){Vernet}, {Fosbury}, {Villar-Mart{\'{\i}}n},
  {Cohen}, {di Serego Alighieri}, \& {Cimatti}}]{VFV99}
{Vernet}, J., {Fosbury}, R.~A.~E., {Villar-Mart{\'{\i}}n}, M., {et~al.} 1999,
  in Astronomical Society of the Pacific Conference Series, Vol. 193, The
  Hy-Redshift Universe: Galaxy Formation and Evolution at High Redshift, ed.
  A.~J. {Bunker} \& W.~J.~M. {van Breugel}, 102

\bibitem[{{V{\'e}ron-Cetty} \& {V{\'e}ron}(2010)}]{VV10}
{V{\'e}ron-Cetty}, M.-P. \& {V{\'e}ron}, P. 2010, \aap, 518, A10

\bibitem[{{Villar-Mart{\'{i}}n} {et~al.}(2003){Villar-Mart{\'{i}}n}, {Vernet},
  {di Serego Alighieri}, {Fosbury}, {Humphrey}, \& {Pentericci}}]{VM03}
{Villar-Mart{\'{i}}n}, M., {Vernet}, J., {di Serego Alighieri}, S., {et~al.}
  2003, \mnras, 346, 273

\bibitem[{{Wardle} \& {Kronberg}(1974)}]{WK74}
{Wardle}, J.~F.~C. \& {Kronberg}, P.~P. 1974, \apj, 194, 249

\bibitem[{{Weidinger} {et~al.}(2005){Weidinger}, {M{\o}ller}, {Fynbo}, \&
  {Thomsen}}]{WMF05}
{Weidinger}, M., {M{\o}ller}, P., {Fynbo}, J.~P.~U., \& {Thomsen}, B. 2005,
  \aap, 436, 825

\bibitem[{{Willott} {et~al.}(2011){Willott}, {Chet}, {Bergeron}, \&
  {Hutchings}}]{WCB11}
{Willott}, C.~J., {Chet}, S., {Bergeron}, J., \& {Hutchings}, J.~B. 2011, \aj,
  142, 186

\bibitem[{{Wisotzki} {et~al.}(2016){Wisotzki}, {Bacon}, {Blaizot},
  {Brinchmann}, {Herenz}, {Schaye}, {Bouch{\'e}}, {Cantalupo}, {Contini},
  {Carollo}, {Caruana}, {Courbot}, {Emsellem}, {Kamann}, {Kerutt}, {Leclercq},
  {Lilly}, {Patr{\'{\i}}cio}, {Sandin}, {Steinmetz}, {Straka}, {Urrutia},
  {Verhamme}, {Weilbacher}, \& {Wendt}}]{WBB16}
{Wisotzki}, L., {Bacon}, R., {Blaizot}, J., {et~al.} 2016, \aap, 587, A98

\bibitem[{{Wright}(2006)}]{W06}
{Wright}, E.~L. 2006, \pasp, 118, 1711

\bibitem[{{Xu} {et~al.}(2022){Xu}, {Cheng}, {Appleton}, {Duc}, {Gao}, {Tang},
  {Yun}, {Dai}, {Huang}, {Lisenfeld}, \& {Renaud}}]{XCA22}
{Xu}, C.~K., {Cheng}, C., {Appleton}, P.~N., {et~al.} 2022, \nat, 610, 461

\bibitem[{{Yang} {et~al.}(2009){Yang}, {Zabludoff}, {Tremonti}, {Eisenstein},
  \& {Dav{\'e}}}]{YZTE09}
{Yang}, Y., {Zabludoff}, A., {Tremonti}, C., {Eisenstein}, D., \& {Dav{\'e}},
  R. 2009, \apj, 693, 1579

\bibitem[{{You} {et~al.}(2017){You}, {Zabludoff}, {Smith}, {Yang}, {Kim},
  {Jannuzi}, {Prescott}, {Matsuda}, \& {Lee}}]{YZS17}
{You}, C., {Zabludoff}, A., {Smith}, P., {et~al.} 2017, \apj, 834, 182

\bibitem[{{Zafar} {et~al.}(2011){Zafar}, {M{\o}ller}, {Ledoux}, {Fynbo},
  {Nilsson}, {Christensen}, {D'Odorico}, {Milvang-Jensen}, {Micha{\l}owski}, \&
  {Ferreira}}]{ZML11}
{Zafar}, T., {M{\o}ller}, P., {Ledoux}, C., {et~al.} 2011, \aap, 532, A51

\bibitem[{{Zirm} {et~al.}(2009){Zirm}, {Dey}, {Dickinson}, \& {Norman}}]{ZDD09}
{Zirm}, A.~W., {Dey}, A., {Dickinson}, M., \& {Norman}, C.~J. 2009, \apjl, 694,
  L31

\end{thebibliography}

\clearpage
\onecolumn

\begin{appendix}
\section{List of sources tested for polarization}

\begin{longtable}{rrrrrrrrrcl}
\caption{\label{tab:pol_all}Polarization of all sources detected at each angle and beam in
all strips recorded by the CHIP1 detector.}\\
\hline
\hline
\multicolumn{2}{c}{Coordinates 
J2000}&$\bar{Q}$&$\bar{U}$&$P$&$P_\mathrm{corr}$&$\sigma(P)$&$\chi$&$\sigma(\chi)$&$P_\mathrm{corr}/\sigma(P)$ & Aspect\\
$\alpha$ (h mn s)&$\delta$ ($\deg$ ' $\arcsec$)&& &         &         &         &($\deg$)  &($\deg$)  &          & \\ \hline
\endfirsthead
\caption{continued}\\
\hline\hline
\multicolumn{2}{c}{Coordinates J2000}&$\bar{Q}$&$\bar{U}$&$P$&$P_\mathrm{corr}$&$\sigma(P)$&$\chi$&$\sigma(\chi)$&$P_\mathrm{corr}/\sigma(P)$ & Aspect\\
$\alpha$ (h mn s)&$\delta$ ($\deg$ ' $\arcsec$)&& &         &         &         &($\deg$)  &($\deg$)  &          & \\ \hline
\endhead
\hline
\endfoot
\hline
\endlastfoot
$12:40:34.93$ & $+14:55:40.6$ & $ 0.01$ & $ 0.04$ & $0.039$ & $0.036$ & $0.016$ & $  39.2$ & $  12.7$ & $   2.3$ & gal.\\
$      34.72$ & $       31.6$ & $ 0.01$ & $ 0.01$ & $0.019$ & $0.019$ & $0.003$ & $  25.9$ & $   5.1$ & $   5.6$ & gal.\\
$      34.38$ & $       37.1$ & $ 0.01$ & $-0.04$ & $0.042$ & $0.039$ & $0.015$ & $ -37.9$ & $  11.2$ & $   2.6$ & star\\
$      32.24$ & $       40.7$ & $-0.02$ & $ 0.08$ & $0.084$ & $0.081$ & $0.025$ & $ -39.4$ & $   8.8$ & $   3.2$ & gal.\\
$      31.30$ & $       43.7$ & $ 0.11$ & $ 0.02$ & $0.112$ & $0.109$ & $0.022$ & $   5.4$ & $   5.8$ & $   4.9$ & star\\
$      29.30$ & $       29.6$ & $ 0.00$ & $-0.01$ & $0.009$ & $0.008$ & $0.004$ & $ -35.9$ & $  15.4$ & $   1.9$ & gal.\\
$      28.31$ & $       43.0$ & $-0.03$ & $-0.07$ & $0.080$ & $0.079$ & $0.015$ & $  34.1$ & $   5.3$ & $   5.4$ & gal.\\
$      27.72$ & $       44.7$ & $-0.02$ & $ 0.01$ & $0.021$ & $0.000$ & $0.024$ & $ -14.7$ & $ 180.0$ & $   0.0$ & gal.\\
$      26.43$ & $       39.7$ & $ 0.01$ & $ 0.00$ & $0.010$ & $0.008$ & $0.005$ & $   2.3$ & $  17.7$ & $   1.6$ & star\\
$      26.24$ & $       27.8$ & $-0.00$ & $-0.01$ & $0.015$ & $0.015$ & $0.004$ & $  37.8$ & $   8.4$ & $   3.4$ & star\\
$      26.13$ & $       39.2$ & $-0.01$ & $-0.09$ & $0.092$ & $0.090$ & $0.020$ & $  40.8$ & $   6.3$ & $   4.5$ & gal.\\
$      25.88$ & $       33.1$ & $-0.03$ & $-0.02$ & $0.031$ & $0.029$ & $0.011$ & $  15.4$ & $  11.2$ & $   2.6$ & gal.\\
$      23.95$ & $       31.1$ & $-0.03$ & $ 0.03$ & $0.042$ & $0.031$ & $0.029$ & $ -24.6$ & $  26.4$ & $   1.1$ & gal.\\
$      23.49$ & $       41.2$ & $-0.00$ & $ 0.14$ & $0.139$ & $0.136$ & $0.031$ & $ -44.8$ & $   6.6$ & $   4.4$ & gal.\\
$      23.25$ & $       37.7$ & $-0.06$ & $ 0.01$ & $0.056$ & $0.046$ & $0.033$ & $  -2.6$ & $  20.9$ & $   1.4$ & star\\
$      22.43$ & $       45.2$ & $-0.05$ & $ 0.01$ & $0.046$ & $0.042$ & $0.019$ & $  -4.4$ & $  13.3$ & $   2.2$ & gal.\\
$      21.74$ & $       29.4$ & $-0.02$ & $ 0.04$ & $0.041$ & $0.033$ & $0.024$ & $ -34.2$ & $  21.2$ & $   1.4$ & star\\
$      20.92$ & $       35.6$ & $-0.00$ & $-0.01$ & $0.007$ & $0.007$ & $0.003$ & $  43.7$ & $  12.1$ & $   2.4$ & QSO \\
$      19.03$ & $       41.7$ & $ 0.12$ & $ 0.12$ & $0.168$ & $0.165$ & $0.030$ & $  23.1$ & $   5.3$ & $   5.5$ & star\\
$      18.36$ & $       26.8$ & $ 0.07$ & $-0.01$ & $0.067$ & $0.061$ & $0.026$ & $  -3.4$ & $  12.2$ & $   2.4$ & star\\
$      16.71$ & $       33.4$ & $ 0.02$ & $-0.07$ & $0.068$ & $0.066$ & $0.020$ & $ -37.3$ & $   8.5$ & $   3.4$ & gal.\\
$      15.62$ & $       45.0$ & $ 0.01$ & $-0.02$ & $0.026$ & $0.025$ & $0.007$ & $ -35.3$ & $   8.4$ & $   3.4$ & gal.\\
$      15.15$ & $       31.6$ & $ 0.04$ & $-0.01$ & $0.038$ & $0.037$ & $0.007$ & $  -3.8$ & $   5.4$ & $   5.3$ & gal.\\
$      14.86$ & $       45.7$ & $ 0.01$ & $-0.01$ & $0.015$ & $0.013$ & $0.007$ & $ -16.0$ & $  16.1$ & $   1.8$ & gal.\\
$      11.43$ & $       36.4$ & $ 0.02$ & $-0.00$ & $0.022$ & $0.021$ & $0.007$ & $  -0.2$ & $  10.3$ & $   2.8$ & gal.\\
$      10.68$ & $       36.2$ & $-0.03$ & $-0.01$ & $0.026$ & $0.016$ & $0.021$ & $   7.9$ & $  37.8$ & $   0.8$ & gal.\\
$      10.07$ & $       40.7$ & $-0.04$ & $ 0.01$ & $0.042$ & $0.035$ & $0.023$ & $  -6.1$ & $  19.4$ & $   1.5$ & gal.\\
$      08.46$ & $       32.1$ & $-0.06$ & $ 0.01$ & $0.061$ & $0.058$ & $0.019$ & $  -2.6$ & $   9.5$ & $   3.0$ & star\\
$      07.85$ & $       30.1$ & $-0.00$ & $-0.02$ & $0.016$ & $0.004$ & $0.016$ & $  43.2$ & $ 118.6$ & $   0.2$ & star\\
$      06.89$ & $       27.8$ & $ 0.01$ & $-0.00$ & $0.009$ & $0.009$ & $0.000$ & $  -2.5$ & $   1.4$ & $  21.0$ & star\\
$      06.65$ & $       39.9$ & $ 0.04$ & $-0.03$ & $0.049$ & $0.045$ & $0.018$ & $ -16.6$ & $  11.7$ & $   2.5$ & star\\
$      06.41$ & $       35.9$ & $ 0.01$ & $-0.00$ & $0.011$ & $0.000$ & $0.026$ & $ -10.2$ & $ 180.0$ & $   0.0$ & gal. \\  \hline
$      33.21$ & $    56:31.6$ & $-0.07$ & $-0.00$ & $0.066$ & $0.062$ & $0.023$ & $   1.4$ & $  10.5$ & $   2.7$ & star\\
$      31.58$ & $       26.0$ & $-0.05$ & $-0.01$ & $0.053$ & $0.052$ & $0.009$ & $   3.1$ & $   4.7$ & $   6.1$ & star\\
$      31.47$ & $       29.5$ & $ 0.01$ & $-0.01$ & $0.015$ & $0.014$ & $0.003$ & $ -27.9$ & $   6.9$ & $   4.1$ & gal.\\
$      31.21$ & $       25.0$ & $ 0.01$ & $ 0.00$ & $0.006$ & $0.006$ & $0.001$ & $   0.4$ & $   6.8$ & $   4.2$ & gal.\\
$      31.05$ & $       31.8$ & $-0.01$ & $ 0.01$ & $0.015$ & $0.012$ & $0.009$ & $ -31.7$ & $  20.8$ & $   1.4$ & gal.\\
$      29.14$ & $       29.6$ & $-0.04$ & $ 0.03$ & $0.055$ & $0.053$ & $0.013$ & $ -17.9$ & $   6.8$ & $   4.2$ & gal.\\
$      28.73$ & $       19.5$ & $ 0.00$ & $-0.00$ & $0.005$ & $0.005$ & $0.001$ & $  -1.9$ & $   6.1$ & $   4.7$ & gal.\\
$      25.18$ & $       26.0$ & $-0.01$ & $-0.01$ & $0.013$ & $0.012$ & $0.005$ & $  24.6$ & $  12.6$ & $   2.3$ & star\\
$      24.60$ & $       20.1$ & $ 0.01$ & $-0.01$ & $0.014$ & $0.012$ & $0.006$ & $ -11.2$ & $  13.4$ & $   2.1$ & star\\
$      22.66$ & $       30.6$ & $-0.02$ & $ 0.09$ & $0.092$ & $0.089$ & $0.025$ & $ -37.3$ & $   8.0$ & $   3.6$ & star\\
$      18.78$ & $       23.8$ & $-0.02$ & $ 0.00$ & $0.020$ & $0.014$ & $0.014$ & $  -0.5$ & $  27.7$ & $   1.0$ & gal.\\
$      17.67$ & $       20.0$ & $ 0.06$ & $-0.10$ & $0.119$ & $0.115$ & $0.029$ & $ -30.2$ & $   7.2$ & $   4.0$ & gal.\\
$      15.82$ & $       27.0$ & $-0.00$ & $-0.03$ & $0.029$ & $0.028$ & $0.007$ & $  41.0$ & $   6.8$ & $   4.2$ & gal.\\
$      14.60$ & $       26.0$ & $ 0.01$ & $ 0.01$ & $0.011$ & $0.010$ & $0.005$ & $  17.2$ & $  16.2$ & $   1.8$ & star\\
$      14.18$ & $       22.0$ & $ 0.06$ & $ 0.00$ & $0.062$ & $0.058$ & $0.022$ & $   1.3$ & $  10.8$ & $   2.6$ & star\\
$      11.95$ & $       21.7$ & $-0.02$ & $-0.04$ & $0.040$ & $0.036$ & $0.018$ & $  33.5$ & $  13.9$ & $   2.1$ & star\\
$      10.65$ & $       22.5$ & $-0.04$ & $ 0.06$ & $0.071$ & $0.066$ & $0.027$ & $ -25.8$ & $  11.6$ & $   2.5$ & gal.\\
$      09.29$ & $       17.7$ & $ 0.05$ & $-0.00$ & $0.047$ & $0.042$ & $0.021$ & $  -1.8$ & $  14.5$ & $   2.0$ & gal.\\
$      08.63$ & $       14.9$ & $-0.00$ & $-0.00$ & $0.003$ & $0.000$ & $0.006$ & $   8.3$ & $ 180.0$ & $   0.0$ & gal.\\
$      08.51$ & $       12.4$ & $-0.12$ & $-0.06$ & $0.130$ & $0.129$ & $0.021$ & $  13.7$ & $   4.6$ & $   6.2$ & star\\
$      06.88$ & $       12.2$ & $ 0.06$ & $ 0.00$ & $0.057$ & $0.054$ & $0.018$ & $   2.3$ & $   9.6$ & $   3.0$ & star \\  \hline
$      34.21$ & $    57:01.2$ & $-0.03$ & $-0.04$ & $0.050$ & $0.049$ & $0.011$ & $  27.6$ & $   6.5$ & $   4.4$ & gal.\\
$      33.09$ & $       08.6$ & $ 0.00$ & $-0.02$ & $0.017$ & $0.016$ & $0.006$ & $ -42.7$ & $   9.7$ & $   3.0$ & gal.\\
$      32.26$ & $    56:58.5$ & $-0.01$ & $ 0.01$ & $0.016$ & $0.014$ & $0.007$ & $ -27.0$ & $  14.4$ & $   2.0$ & star\\
$      31.96$ & $    57:09.4$ & $-0.00$ & $-0.02$ & $0.019$ & $0.016$ & $0.010$ & $  41.3$ & $  17.4$ & $   1.6$ & gal.\\
$      30.34$ & $    57:15.5$ & $-0.00$ & $ 0.01$ & $0.007$ & $0.000$ & $0.007$ & $ -26.4$ & $ 180.0$ & $   0.0$ & gal.\\
$      27.75$ & $    56:59.0$ & $ 0.15$ & $-0.04$ & $0.154$ & $0.152$ & $0.021$ & $  -6.6$ & $   4.0$ & $   7.1$ & star\\
$      24.52$ & $    57:02.8$ & $-0.03$ & $-0.07$ & $0.077$ & $0.076$ & $0.015$ & $  33.1$ & $   5.5$ & $   5.2$ & gal.\\
$      24.35$ & $    57:09.6$ & $ 0.09$ & $ 0.00$ & $0.088$ & $0.086$ & $0.019$ & $   0.3$ & $   6.4$ & $   4.5$ & gal.\\
$      21.67$ & $    57:04.3$ & $-0.03$ & $ 0.01$ & $0.033$ & $0.027$ & $0.018$ & $ -12.7$ & $  19.4$ & $   1.5$ & gal.\\
$      20.82$ & $    57:03.5$ & $-0.02$ & $-0.02$ & $0.029$ & $0.028$ & $0.010$ & $  18.7$ & $  10.4$ & $   2.8$ & star\\
$      18.68$ & $    57:03.3$ & $-0.00$ & $ 0.00$ & $0.002$ & $0.001$ & $0.001$ & $ -29.4$ & $  20.8$ & $   1.4$ & gal\\
$      17.65$ & $    57:13.9$ & $-0.02$ & $ 0.00$ & $0.016$ & $0.015$ & $0.006$ & $  -5.9$ & $  11.4$ & $   2.5$ & star\\
$      17.50$ & $    57:01.0$ & $ 0.06$ & $ 0.03$ & $0.067$ & $0.064$ & $0.017$ & $  11.2$ & $   7.7$ & $   3.7$ & gal.\\
$      15.53$ & $    56:59.3$ & $-0.04$ & $ 0.03$ & $0.050$ & $0.048$ & $0.014$ & $ -19.9$ & $   8.5$ & $   3.4$ & gal.\\
$      14.61$ & $    56:57.8$ & $ 0.00$ & $ 0.00$ & $0.004$ & $0.003$ & $0.001$ & $  44.7$ & $   6.8$ & $   4.2$ & gal.\\
$      14.11$ & $    57:00.0$ & $-0.05$ & $-0.06$ & $0.076$ & $0.073$ & $0.020$ & $  24.6$ & $   7.8$ & $   3.7$ & gal.\\
$      12.42$ & $    56:57.1$ & $ 0.04$ & $ 0.04$ & $0.055$ & $0.050$ & $0.022$ & $  24.3$ & $  12.6$ & $   2.3$ & star\\
$      11.10$ & $    57:00.8$ & $ 0.00$ & $ 0.01$ & $0.014$ & $0.013$ & $0.006$ & $  42.5$ & $  12.7$ & $   2.3$ & gal.\\
$      10.42$ & $    57:03.1$ & $ 0.04$ & $ 0.00$ & $0.039$ & $0.033$ & $0.022$ & $   2.3$ & $  19.4$ & $   1.5$ & gal.\\
$      08.96$ & $    57:05.3$ & $ 0.04$ & $-0.02$ & $0.041$ & $0.041$ & $0.005$ & $ -13.7$ & $   3.4$ & $   8.5$ & gal.\\
$      07.02$ & $    57:07.1$ & $-0.04$ & $ 0.06$ & $0.075$ & $0.072$ & $0.021$ & $ -27.4$ & $   8.5$ & $   3.4$ & gal. \\  \hline
$      34.29$ & $    57:59.2$ & $ 0.00$ & $-0.04$ & $0.042$ & $0.042$ & $0.006$ & $ -44.4$ & $   4.2$ & $   6.9$ & star\\
$      34.22$ & $       46.1$ & $-0.03$ & $-0.12$ & $0.122$ & $0.121$ & $0.014$ & $  37.6$ & $   3.2$ & $   8.8$ & gal.\\
$      33.13$ & $       58.0$ & $ 0.02$ & $ 0.05$ & $0.052$ & $0.048$ & $0.020$ & $  32.8$ & $  11.9$ & $   2.4$ & gal.\\
$      32.00$ & $       57.2$ & $-0.00$ & $-0.01$ & $0.007$ & $0.005$ & $0.005$ & $  37.7$ & $  29.8$ & $   1.0$ & gal.\\
$      26.64$ & $       55.7$ & $-0.09$ & $ 0.02$ & $0.096$ & $0.093$ & $0.021$ & $  -6.2$ & $   6.5$ & $   4.4$ & gal.\\
$      26.12$ & $       52.4$ & $-0.01$ & $-0.01$ & $0.016$ & $0.012$ & $0.010$ & $  18.2$ & $  22.3$ & $   1.3$ & gal.\\
$      22.43$ & $       44.1$ & $-0.04$ & $-0.02$ & $0.042$ & $0.040$ & $0.013$ & $  14.8$ & $   9.2$ & $   3.1$ & star\\
$      20.05$ & $       44.1$ & $ 0.02$ & $-0.02$ & $0.032$ & $0.030$ & $0.012$ & $ -22.9$ & $  11.1$ & $   2.6$ & gal.\\
$      19.44$ & $       59.8$ & $-0.02$ & $-0.00$ & $0.025$ & $0.023$ & $0.008$ & $   1.7$ & $  10.3$ & $   2.8$ & star\\
$      19.17$ & $       44.4$ & $ 0.02$ & $ 0.03$ & $0.034$ & $0.023$ & $0.025$ & $  26.7$ & $  31.5$ & $   0.9$ & star\\
$      18.81$ & $       57.7$ & $-0.01$ & $ 0.03$ & $0.028$ & $0.026$ & $0.009$ & $ -35.4$ & $  10.0$ & $   2.9$ & gal.\\
$      15.79$ & $       46.7$ & $-0.00$ & $ 0.00$ & $0.004$ & $0.004$ & $0.001$ & $ -25.2$ & $   8.2$ & $   3.5$ & star\\
$      15.79$ & $       57.8$ & $ 0.02$ & $-0.02$ & $0.029$ & $0.024$ & $0.016$ & $ -23.4$ & $  18.9$ & $   1.5$ & gal.\\
$      15.60$ & $    58:00.0$ & $ 0.01$ & $-0.00$ & $0.012$ & $0.011$ & $0.004$ & $  -3.8$ & $  11.1$ & $   2.6$ & star\\
$      14.40$ & $    57:58.0$ & $-0.03$ & $-0.06$ & $0.067$ & $0.060$ & $0.028$ & $  32.6$ & $  13.3$ & $   2.1$ & star\\
$      13.96$ & $    57:59.0$ & $ 0.04$ & $-0.05$ & $0.070$ & $0.064$ & $0.029$ & $ -25.5$ & $  13.1$ & $   2.2$ & star\\
$      12.58$ & $    57:58.7$ & $ 0.02$ & $-0.05$ & $0.051$ & $0.049$ & $0.013$ & $ -35.4$ & $   7.7$ & $   3.7$ & gal.\\
$      11.92$ & $       44.9$ & $ 0.08$ & $-0.01$ & $0.085$ & $0.084$ & $0.013$ & $  -3.1$ & $   4.4$ & $   6.5$ & gal.\\
$      07.43$ & $    58:00.2$ & $ 0.03$ & $ 0.02$ & $0.036$ & $0.034$ & $0.013$ & $  13.2$ & $  11.2$ & $   2.6$ & gal. \\  \hline
$      31.14$ & $    58:39.0$ & $ 0.02$ & $ 0.01$ & $0.025$ & $0.012$ & $0.023$ & $   6.1$ & $  55.2$ & $   0.5$ & gal.\\
$      29.98$ & $       43.8$ & $-0.02$ & $-0.02$ & $0.028$ & $0.024$ & $0.016$ & $  20.2$ & $  18.7$ & $   1.5$ & gal.\\
$      29.32$ & $       35.5$ & $ 0.02$ & $-0.07$ & $0.068$ & $0.061$ & $0.030$ & $ -36.9$ & $  14.4$ & $   2.0$ & gal.\\
$      27.02$ & $       41.3$ & $-0.05$ & $ 0.01$ & $0.051$ & $0.049$ & $0.015$ & $  -4.4$ & $   8.6$ & $   3.3$ & gal.\\
$      26.75$ & $       40.6$ & $ 0.01$ & $ 0.08$ & $0.078$ & $0.071$ & $0.031$ & $  40.9$ & $  12.4$ & $   2.3$ & gal.\\
$      23.46$ & $       37.3$ & $-0.02$ & $-0.01$ & $0.020$ & $0.013$ & $0.016$ & $   8.4$ & $  35.8$ & $   0.8$ & gal.\\
$      21.46$ & $       29.7$ & $-0.05$ & $-0.01$ & $0.052$ & $0.047$ & $0.023$ & $   7.3$ & $  14.1$ & $   2.0$ & gal.\\
$      17.93$ & $       37.0$ & $ 0.00$ & $-0.06$ & $0.056$ & $0.048$ & $0.028$ & $ -42.5$ & $  16.6$ & $   1.7$ & star\\
$      17.39$ & $       48.1$ & $-0.03$ & $ 0.01$ & $0.033$ & $0.031$ & $0.012$ & $  -9.5$ & $  11.3$ & $   2.5$ & star\\
$      16.82$ & $       29.0$ & $ 0.00$ & $ 0.00$ & $0.006$ & $0.000$ & $0.015$ & $  15.8$ & $ 180.0$ & $   0.0$ & star\\
$      14.35$ & $       34.0$ & $ 0.03$ & $ 0.02$ & $0.037$ & $0.026$ & $0.026$ & $  12.1$ & $  28.4$ & $   1.0$ & star\\
$      11.90$ & $       35.0$ & $ 0.01$ & $ 0.05$ & $0.052$ & $0.049$ & $0.017$ & $  38.8$ & $   9.8$ & $   2.9$ & gal.\\
$      10.42$ & $       37.8$ & $-0.02$ & $ 0.07$ & $0.072$ & $0.070$ & $0.015$ & $ -37.8$ & $   6.2$ & $   4.6$ & gal.\\
$      10.02$ & $       36.7$ & $-0.03$ & $ 0.05$ & $0.058$ & $0.056$ & $0.013$ & $ -30.0$ & $   6.7$ & $   4.3$ & gal.\\
$      09.73$ & $       31.0$ & $ 0.06$ & $ 0.11$ & $0.131$ & $0.127$ & $0.029$ & $  30.3$ & $   6.4$ & $   4.5$ & gal. \\  \hline
\end{longtable}
\tablefoot{ We give the equatorial coordinates,
measured $\bar{Q}$, $\bar{U}$, raw and corrected polarization fractions $P$
(see Sect.\,\ref{sec:polar}) and their error, angle $\chi$ (corrected for
instrumental effect, see Subsect.\,\ref{subsec:polcal}) and its error,
and signal to noise ratio of $P$. The last column indicates the appearance of
the object, point source ("star") or extended object ("gal."). When $P_\mathrm{corr}=0$,
the angle $\chi$ is actually undefined and its error is set to $180\deg$; likewise,
the signal to noise on $P_\mathrm{corr}$ is set to zero.
The five blocks represent the successive $20$\arcsec strips that are arranged
from south to north.}
\end{appendix}
\end{document}